\documentclass[aps,prx, twocolumn, floatfix, longbibliography]{revtex4-2}

\usepackage[utf8]{inputenc}
\usepackage{graphicx}
\usepackage{wrapfig}
\usepackage{mathtools}
\usepackage{setspace}
\usepackage{xcolor}
\usepackage{multirow}
\graphicspath{ {Figures/} }

\usepackage{soul}

\newcommand{\eqnref}[1]{eqn.~(\ref{#1})}

\newcommand{\figref}[1]{fig.~\ref{#1}}
\newcommand{\V}[1]{\mathbf{#1}}
\newcommand{\Vh}[1]{\mathbf{\hat #1}}

\begin{document}
\begin{titlepage}
   \begin{center}

       \textbf{\huge Dynamics of Acoustically Bound Particles}
            
       \vspace{1.0cm}

       \textbf{Nicholas St. Clair$^1$, Dominique Davenport$^1$, Arnold D. Kim$^2$, Dustin Kleckner$^{1,*}$}
       
       \vspace{0.5cm}
       
       Department of Physics$^1$, Department of Applied Mathematics$^2$\\
       School of Natural Sciences, University of California, Merced\\
       United States\\
       February 2022
       
       \vspace{0.9cm}
    \textbf{Abstract}
    \end{center}

It is well known that scattering from acoustic fields can produce forces on single particles, however they can also induce interparticle forces due to multiple scattering events.
This multi-particle force -- here referred to as acoustic binding -- is comparable to other acoustic forces when the particles are of order wavelength in diameter.
In principle, this force could be used as a tunable method for directing the assembly of particles of mm-scales, but has not been extensively explored in previous work.
Here, we use a novel numerical method to compute binding interactions between strongly scattering bodies and find that they can produce stable clusters of particles with approximately wavelength separation.
Moreover, we also observe that -- depending on the level of damping -- these structures can produce driven linear, rotational, or vibrational motion.
These effects are a result of the non-conservative and non-pairwise nature of the acoustic binding force, and represent novel contactless manipulation and transport methods with a variety of potential applications. 

    \vfill

    \vspace{0.8cm}
        
\end{titlepage}

\section{Introduction}

While matter can alter the path of an external field through scattering and absorption, the field can also induce forces on matter.
The resulting forces are a topic of great interest within active and condensed matter communities, as well as the field of material synthesis \cite{perryTwoDimensionalClustersColloidal2015, mengFreeEnergyLandscapeClusters2010, lenshofAcoustofluidicsApplicationsAcoustophoresis2012}.
For micron-scale objects, optical trapping has found numerous applications in physics, biology, and medicine \cite{neumanOpticalTrapping2004, ashkinAccelerationTrappingParticles1970, maragoOpticalTrappingManipulation2013, ashkinOpticalTrappingManipulation1987, ashokOpticalTrappingAnalytical2012, pangOpticalTrappingSingle2012, woerdemannAdvancedOpticalTrapping2013}.
Optical forces can be placed on single objects, however second-order effects, known as optical binding can introduce inter-particle forces that has the potential to self-organize structures with wavelength-scale features \cite{burnsOpticalBinding1989b, ngPhotonicClustersFormed2005}.
These binding forces arise when scattered and incident fields interfere to produce a new field gradient between two or more scattering bodies.
\begin{figure}
\includegraphics[width = 0.95\linewidth]{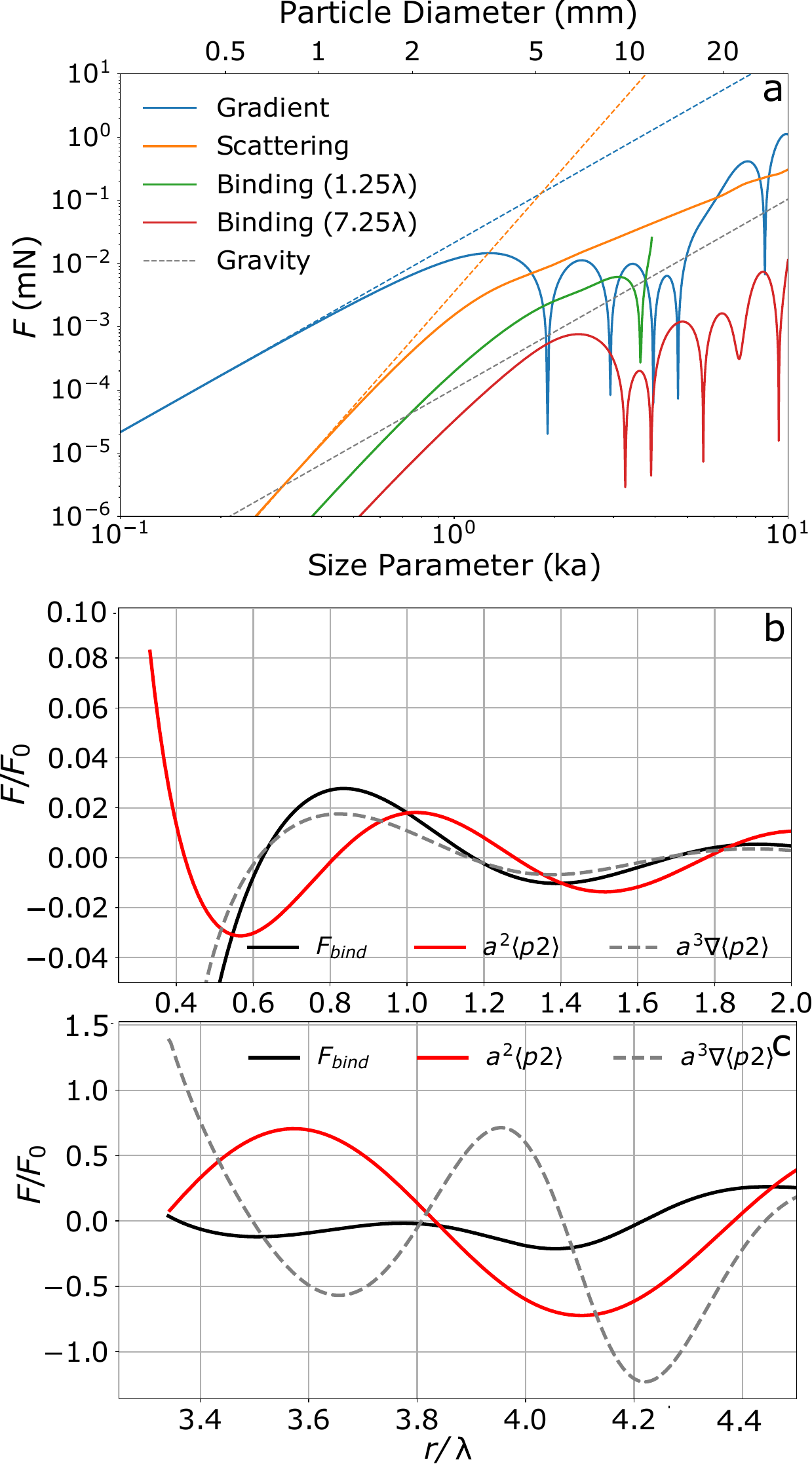}
\caption{Acoustic force scaling.
{\bf a)} Relative strengths of various acoustic forces as a function of size of the scattering body, computed using MFS numerical model. 
For $ka \lesssim 1$ the gradient force (blue) dominates; for $ka \gtrsim 1$ the secondary forces (red, green, orange) are of comparable magnitude.
The binding force is shown for two particles at fixed separation distances of $1.25 \lambda$ (green) and $7.25\lambda$ (red) while the gradient force is for a single particle located between a pressure node and antinode. For reference the particle volume (dashed blue) and square of the volume (dashed orange) are included.
{\bf b,c)} Two particle acoustic binding force (black), 2nd order pressure field ($\langle p_2 \rangle$) (red), and gradient of the 2nd order pressure field integrated over particle surface (dashed gray) vs particle separation for {\bf(b)} $ka=1$  and {\bf(c)} $ka=10$.
\label{fig:Sizes}}
\end{figure}

Could optical forces be used to manipulate larger particles?
The field carried momentum goes as the intensity over the velocity of the wave $p=I/c$, which restricts the size of bodies that can be manipulated by an external field.
Current optical trapping methods can levitate particles on the order of 10 $\mu$m in size (10 ng in mass) \cite{mooreSearchingNewPhysics2021}, limited by the maximum practical power density.
To manipulate larger particles we can employ a wave with slower speed, such as sound, which produces a force which is roughly $10^6$ times higher for the same level of input power.
Moreover, sound waves have a longer wavelength, so the analogous `acoustic binding' effects will also self-organize structure on larger scales.

Ultrasound has previously been demonstrated to produce forces on single particles or even between many particles \cite{bruusAcoustofluidicsAcousticRadiation2012b, rabaudAcousticallyBoundMicrofluidic2011, limClusterFormationAcoustic2019, jiaoExperimentalTheoreticalStudies2013, peterssonFreeFlowAcoustophoresis2007, sepehrirahnamaNumericalStudyInterparticle2015, courtneyDexterousManipulationMicroparticles2013, baaschAcoustofluidicParticleDynamics2018, doinikovAcousticRadiationInterparticle2001, zhengAcousticInteractionForces1994}.
Acoustic levitation and trapping have been widely used to manipulate single particles on millimeter scales \cite{wuAcousticalTweezers1991, wangDextrousAcousticTrapping2015, abdelazizDynamicsAcousticallyTrapped2021, marzoHolographicAcousticElements2015a, bareschObservationSingleBeamGradient2016, leeSingleBeamAcoustic2009, urbanskyRapidEffectiveEnrichment2017, jonssonParticleSeparationUsing2004}.
(Commonly used 40 kHz sound waves have a wavelength of about 8.6 mm.)
Far less known is the corresponding multiple particle force; in previous literature this has gone by several names -- including acoustically induced mutual force \cite{wangSoundmediatedStableConfigurations2017a}
and acoustic interaction force \cite{lopesAcousticInteractionForces2016}, 
but here we will refer to this force as acoustic binding.
The acoustic binding force arises from interference between the scattered field and the incident, resulting a long range oscillatory force.
As discussed below, this force is distinct from the secondary Bjerknes force, which is a short range interparticle force acting on deformable bodies like bubbles in acoustic fields \cite{yoshidaExperimentalInvestigationReversal2011, castroDeterminationSecondaryBjerknes2015, zhangSecondaryBjerknesForce2016}.

Unlike its optical counterpart, the acoustic binding force has only been studied in a small number of specific cases \cite{sepehrirahnamaNumericalStudyInterparticle2015, garcia-sabateExperimentalStudyInterparticle2014, baaschMultibodyDynamicsAcoustophoresis2017, lopesAcousticInteractionForces2016, wangSoundmediatedStableConfigurations2017a}, 
despite the fact it represents a potentially powerful tool for self-organizing structures on mm-scales.
\begin{figure*}
\begin{centering}
\includegraphics[width = 0.95\linewidth]{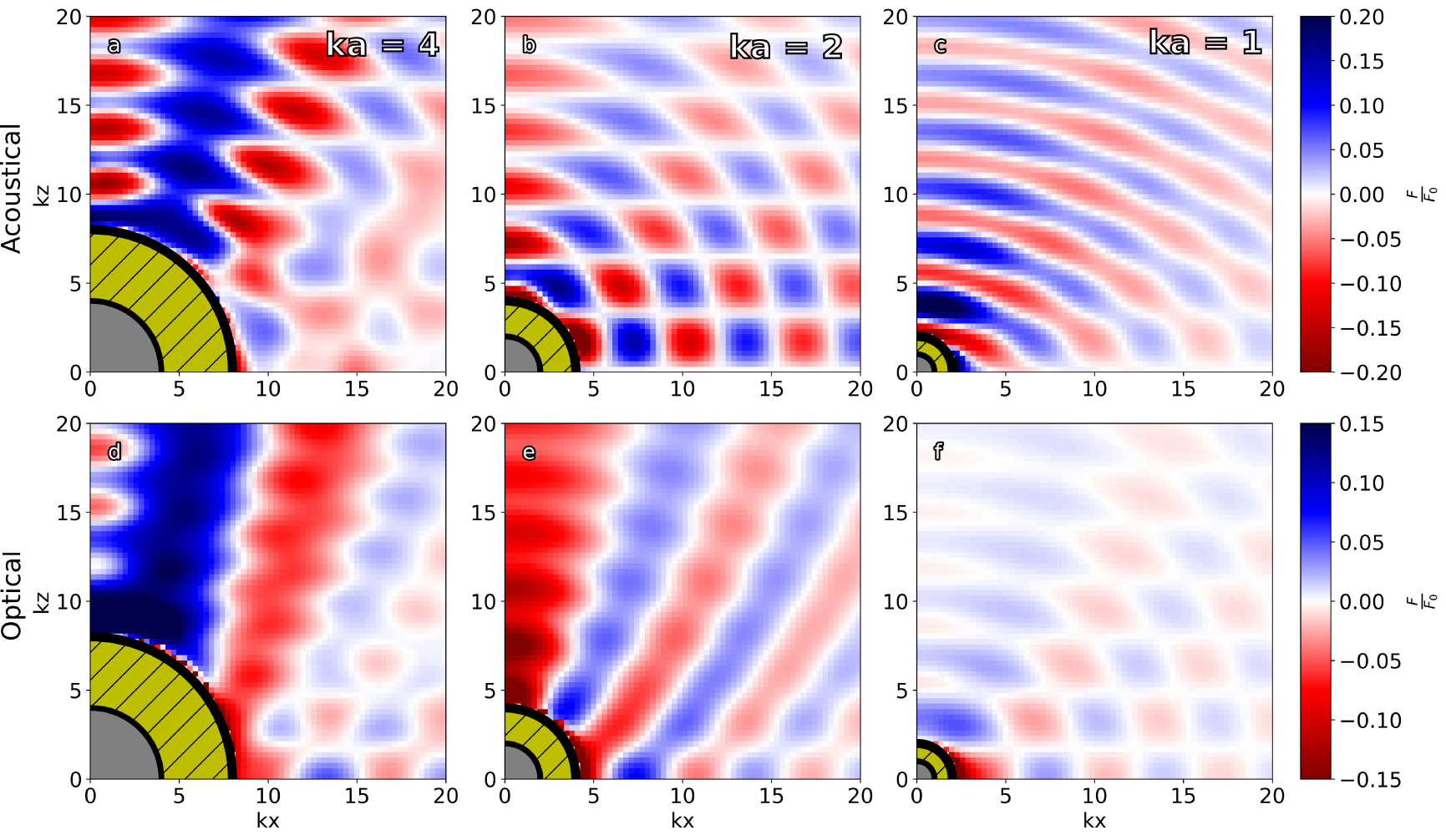}
\end{centering}
\caption{ Acoustic vs.~optical binding. Force maps for acoustic binding ({\bf{a-c}}) and optical binding ({\bf d-f}) for different sizes of particles, $ka = 1$ ({\bf c, f}), 2 ({\bf b, e}), and 4 ({\bf a, d}). 
Both maps are created by keeping one particle (gray quarter-sphere) fixed at the origin, which here corresponds to a location in the incident field halfway between node and antinode, and computing the scattered fields (and hence total force) on a second particle centered at a given location on the map. The single-particle force is subtracted from the computed total force at each location so that these maps show only the radial binding force between two identical particles.
For the acoustic maps the MFS code was implemented, see supplemental materials section E for a discussion of the simulation of optical forces.
The forces in these plots are all scaled to the corresponding acoustical/optical reference force. The hashed yellow regions correspond to exclusion zones within which the two particles would overlap.
\label{fig:Maps}
}
\end{figure*}
Several different methods have been used to model acoustic forces, including multipole expansion of the acoustic scattering from weak scatterers \cite{garcia-sabateExperimentalStudyInterparticle2014, silvaAcousticInteractionForces2014, doinikovAcousticRadiationInterparticle2001}, finite element methods \cite{baaschMultibodyDynamicsAcoustophoresis2017, glynne-jonesEfficientFiniteElement2013}, and a simplification of the problem to pairwise interactions \cite{zhangAcousticallyMediatedLongrange2016}.
Here, we explore the acoustic binding forces in more detail using a novel numerical method, known as the method of fundamental solutions (MFS) \cite{fairweatherMethodFundamentalSolutions2003a}.
This method is both fast and accurate, allowing for one to conduct molecular dynamics simulations of small particle clusters.

For particles in the Mie scattering regime ($ka \gtrsim 1$, where $k$ is the wavenumber of the ultrasound field and $a$ is the particle radius) we find that binding forces become comparable to trapping forces (\figref{fig:Sizes}).
Consequently, these forces can be used to self-assemble structure on wavelength scales, and it should be possible to alter this structure by changing the properties of the particles or acoustic field.
Furthermore, we observe active clusters of particles which drift, oscillate, and/or rotate depending on the cluster configuration (for example, see Supplementary Movies 1, 2, and 3).
As a result, acoustically bound clusters are not limited to passive self-assembly, opening up the possibility of driven particle assembly analogous to recently developed active matter systems \cite{ramaswamyActiveMatter2017,kangRecentProgressActive2019}.
Previous examples of acoustically driven systems have relied on body asymmetry \cite{zhouTwistsTurnsOrbiting2017a}, complex sound fields  \cite{abdelazizDynamicsAcousticallyTrapped2021, mitriAxialAcousticRadiation2012}, or acoustohydrodynamic interactions \cite{abdelazizUltrasonicChainingEmulsion2021, liuInvestigationEffectAcoustic2017, wuAcousticStreamingIts2018} as the driving mechanism.
Conversely the behavior demonstrated here is not a result of any complex particle or beam characteristics, but rather the result of the nonlinearity of acoustic binding interactions between the constituents of this simple system.

\subsection{Acoustic Forces}

Acoustic fields can induce many types of forces on individual particles or between them.
Broadly speaking, these can be separated into scattering forces \cite{bruusAcoustofluidicsAcousticRadiation2012b,doinikovAcousticRadiationInterparticle2001, wangSoundmediatedStableConfigurations2017a}, those due to streaming induced by acoustic waves \cite{liuInvestigationEffectAcoustic2017, wuAcousticStreamingIts2018}, and those due to the deformation of the particles (e.g.~Bjerknes forces) \cite{zhangSecondaryBjerknesForce2016, castroDeterminationSecondaryBjerknes2015,rabaudAcousticallyBoundMicrofluidic2011}.
As detailed below, for wavelength-sized (Mie) solid particles in a gaseous medium, the scattering forces will dominate over these other types.
The scattering forces can be further divided into gradient forces, radiation pressure forces, and binding forces; each of these is directly analogous to the equivalent optical forces \cite{pangOpticalTrappingSingle2012, ashkinOpticalTrappingManipulation1987, ashokOpticalTrappingAnalytical2012, burnsOpticalBinding1989b, neumanOpticalTrapping2004}.
For small particles ($ka \ll 1$), the acoustic gradient force is given by \cite{bruusAcoustofluidicsAcousticRadiation2012b}:
\begin{align}
    \V F_{g} &= - \nabla U_{rad}\\
    U_{rad} &= \frac{4 \pi}{3} a^{3} \left[ f_{1} \frac{1}{2} \kappa_{0} \langle p^{2}_{in} \rangle - f_{2} \frac{3}{4} \rho_{0} \langle v^{2}_{in} \rangle \right]\\
    f_{1} &= 1 - \frac{\kappa_{p}}{\kappa_{0}}\\
    f_{2} &= \frac{2\left(\frac{\rho_{p}}{\rho_{0}}-1\right)}{2\frac{\rho_{p}}{\rho_{0}}+1}
\end{align}
where $\rho$ represents the mass density, $\kappa=(\rho c^{2})^{-1}$ is the compressibility, and $c$ denotes the sound speed; subscripts $p$ and $0$ refer to properties of the scattering particle and background medium, respectively.
Angle brackets denote a time average over one oscillation period (i.e.~$\left< g \right> = \frac{1}{T} \int_{0}^{T}g\ dt$).
In general, this force will tend to pull particles into pressure maxima or minima, depending on the particle properties.

In this manuscript we consider `sound-hard' particles, for which the particle density is much greater than the background medium and the particle compressibilty is much less than the background medium.  
In this approximation -- which is quite good for solid particles in a gaseous medium -- $f_1 = f_2 = 1$.
Conceptually, the gradient force can be regarded as being caused by a single scattering of the acoustic wave from the particle, and consequently it scales like the particle volume.
For larger particles ($ka \gtrsim 1$), this force must be computed numerically from the scattered field.

Other forces arise due to higher order scattering events.
We will separate these into the radiation pressure force, which acts on a single particle, and the acoustic binding force, which acts on clusters of two or more particles.
For small particles ($ka \ll 1$), both of these effects scale like the \emph{square} of the particle volume (\figref{fig:Sizes}), and as a result are only comparable in magnitude to the gradient force when the particle is of order wavelength in size.
The radiation pressure force pushes the particle in the propagation direction of the acoustic wave; this is often ignored because it cancels in counter-propagating beams (e.g.~as used in a conventional acoustic levitation setup). 

The acoustic binding force arises due to the interference of the external field and the field scattered from a neighboring particle. 
Because interference modifies the total field intensity in the vicinity of the particle, nearby particles feel an additional force due to this modulation. 
For small particles, this can be approximated as the gradient of the total field (\figref{fig:Sizes}\textbf{b}), but this approximation breaks down in the Mie regime (\figref{fig:Sizes}\textbf{c}).
The binding force can be particularly difficult to model when particles are close together due to strong multiple scattering events.

Other acoustic forces can arise due to acoustic streaming or deformation of the particles; each of these effects can also induce interparticle forces under the right circumstances.
However, both of these effects are negligible for wavelength sized solid particles in air.

Acoustic streaming is a non-zero mean flow induced by an acoustic field \cite{liuInvestigationEffectAcoustic2017}, which subsequently induces a drag force on any particles embedded in this flow \cite{wuAcousticStreamingIts2018, liuInvestigationEffectAcoustic2017, wiklundAcoustofluidics14Applications2012}.
One can estimate a critical particle radius below which streaming dominates over other second order acoustic effects \cite{bruusAcoustofluidics10Scaling2012a}:
\begin{align}
    a_c = \sqrt{\frac{4 \nu \Psi}{3 \omega \Phi}}
\end{align}
This threshold depends on the kinematic viscosity of the fluid ($\nu$), the sound frequency ($\omega$), acoustophoretic contrast factor ($\Phi$), and a factor that depends on the geometry of the stream generating surface ($\Psi$) which is of order unity.
For 40 kHz ultrasonic fields in air this critical radius is around 10$\mu$m, and so the scattering forces dominate for the Mie sized ($a \approx$ mm) bodies investigated here.

Bjerknes forces result from deformation of `particles' in the acoustic field, giving rise to a non-zero time averaged pressure on the particle surface, and are generally significant only for situations where the particle compressibilty is the same or greater than the background medium (e.g.~bubbles in water) \cite{zhangSecondaryBjerknesForce2016, castroDeterminationSecondaryBjerknes2015}.
For the sound-hard approximation made here, this force is identically zero, however for solid particles in a gaseous medium it will be orders of magnitude smaller than scattering induced forces.

Finally, we note that in this manuscript we do not consider acoustically mediated torques which may cause \textit{individual} bodies to spin. 
The modelled incident field (planar standing wave) carries no angular momentum, as is generally the case for the fields in conventional acoustic trapping devices \cite{lopesAcousticInteractionForces2016, zhangAcousticallyMediatedLongrange2016, garcia-sabateExperimentalStudyInterparticle2014, wangSoundmediatedStableConfigurations2017a, leeSingleBeamAcoustic2009}.
Scattering between three or more bodies can yield asymmetries in the total sound field at the surface of these bodies, however unless they are sound absorbing an interaction torque will not be imparted \cite{lopesAcousticInteractionForces2016, mitriAxialAcousticRadiation2012, silvaRadiationTorqueProduced2012}.

For the remainder of this manuscript then we will consider only forces generated due to sound scattering events.
The scaling of these three forces (gradient, scattering, and binding) can be computed analytically for particles in the Rayleigh regime \cite{bruusAcoustofluidicsAcousticRadiation2012b, silvaAcousticInteractionForces2014}, but this quickly becomes intractable for larger particles.
This is especially problematic as this is precisely the regime in which the binding force becomes comparable to the gradient force.
Although approximate forms exist for very weakly scattering or widely separated particles \cite{zhangAcousticallyMediatedLongrange2016}, such approximations are not valid for realistic experimental conditions.
This is especially true if one wishes to use acoustic forces to guide the self-organization of large particle assemblies, in which case we expect many particles to be in physical contact.

\section{Numerical Methods}

We model the acoustic field using a complex oscillating potential field, $\phi(\V r) \propto e^{-i\omega t}$, which is related to first-order velocity, pressure, and density fluctuations in the following way \cite{bruusAcoustofluidicsPerturbationTheory2012}: 
\begin{align}
\V v_1(\V r) &= \nabla \phi (\V r) \label{eq:velocity}\\
p_1(\V r) &= p_0 + i \rho_{0} \omega \phi (\V r) \label{eq:pressure}\\
\rho_1(\V r) &= \rho_0 \left[ 1 + i \frac{\omega}{c^2} \phi (\V r)\right] \label{eq:density},
\end{align}
where $\omega$ is the sound frequency, $p$ is the pressure, $\rho$ is the density, and $\V v$ is the local fluid velocity.
In all cases, the physical fields are given by the real part of the complex representation.
\begin{figure*}
\begin{centering}
\includegraphics[width = 0.95\linewidth]{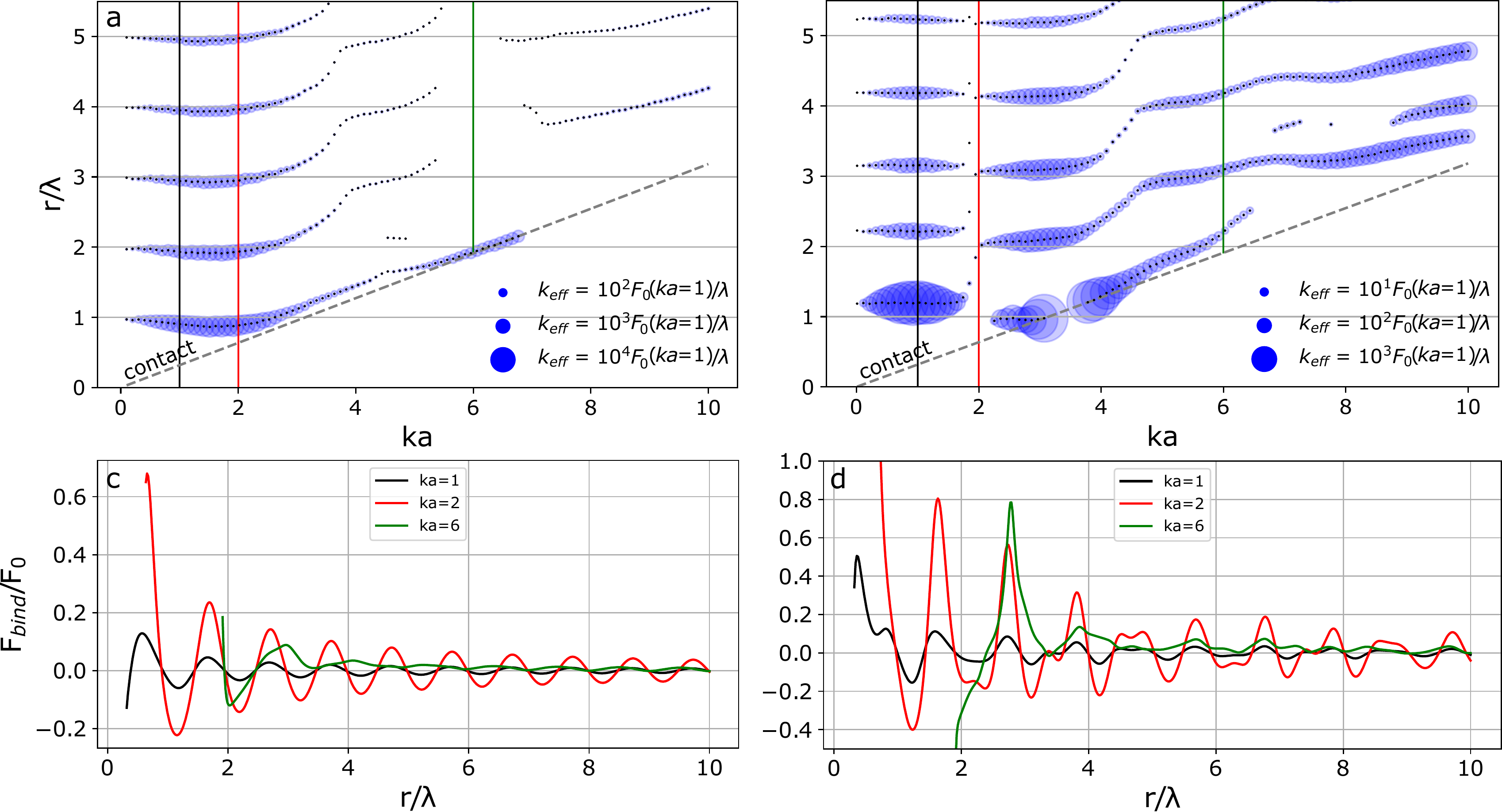}
\end{centering}
\caption{
{{\bf a, b)} First five stable separation distances for two particles side-by-side ({\bf a}) and for a hexagonal cluster of seven particles ({\bf b}).
The size of the encompassing blue marker at each distance encodes the strength of the relative spring constant associated with that equilibrium location. 
The effective spring constants are de-dimensionalized according to the acoustic wavelength and corresponding reference force for a given particle size.
Bottom: Acoustic binding force between two ({\bf c}) and seven ({\bf d}) particles of size parameter $ka=$1 (black), 2 (red), and 6 (green), shown as vertical lines of the corresponding color on {\bf a, b}. 
For the studies involving seven particles, the cluster was arranged in a regular hexagonal pattern and the force was computed on the right-most particle in the cluster.} 
\label{fig:keff}}
\end{figure*}
This linear wave field produces zero net force on an embedded particle if averaged over a full cycle, however, such a formulation does not satisfy the Navier-Stokes equation without including higher order terms.
A perturbation approximation gives rise to second order corrections  -- $p_2$, $\rho_2$, and $\V v_2$ -- which can be formulated in terms of the first order terms.
These second order terms invoke a nonzero net force when averaged over a cycle of the wave \cite{bruusAcoustofluidicsAcousticRadiation2012b}:
\begin{align}
    \left< \V F \right> &=-\int da \left[\langle p_{2}\rangle \hat{n}+\rho_{0}\langle (\Vh n \cdot \V v_{1}) \V v_{1}\rangle \right] \label{eq:Tensor} \\
    \langle p_{2} \rangle &= \frac{1}{4} \left[\kappa_{0} |p_{1}^{2}| - \rho_0 |v_{1}^{2}| \right]  \label{eq:p2},
\end{align}
where the integral in the force calculation is taken over the surface of each particle, and $\phi$ represents the \emph{total} potential field, including both external and scattering contributions. 
Here we have assumed that $p_1 \ll p_0$, in which case higher order terms ($p_3$, etc.) are negligible \cite{bruusAcoustofluidicsPerturbationTheory2012}.
In this work we are interested in the case of solid particles in a gaseous medium.
Using the sound-hard approximation, this is equivalent to $\V v_1 \cdot \Vh n = 0$, removing the second term in \eqnref{eq:Tensor}.
(See Supplemental Materials section D for a discussion of the validity of this approximation.)

We solve for the total first order field, $\phi$, including an external driving input and multiple scattering effects from more than one particle, using the method of fundamental solutions \cite{fairweatherMethodFundamentalSolutions2003a}.
In this model, the scattered field outside the particles is computed from a lattice of virtual scatterers placed inside each particle.
We solve for the amplitude of these virtual scatterers by enforcing the sound-hard boundary condition on another lattice of evaluation points located on the surface of each particle.
In practice, this is done by casting the problem as a linear matrix equation and solved using standard numerical techniques.
The number of points in each lattice affects model accuracy at the cost of computation time; the larger the scattering body, the greater number of source and evaluation points are required to fully resolve the local field oscillations on the surface of that body.
Since the method of fundamental solutions is a surface-based scattering approach, it requires less computational resources than volume-based scattering approaches such as FDTD, DDA, and FEM \cite{glynne-jonesEfficientFiniteElement2013, liuInvestigationEffectAcoustic2017, yurkinDiscreteDipoleApproximation2007a, yurkinDiscretedipoleapproximationCodeADDA2011a}.
Once the total field has been solved, it is numerically integrated over the surface of each particle using a Gaussian quadrature rule \cite{atkinsonNumericalIntegrationSphere1982} to obtain a per-particle force.
More details on the numerical method, and an analysis of its accuracy, are provided in Supplemental Materials section B.

\section{Results}

\subsection{Acoustic Force Scaling}
The relative strength of binding, scattering, and gradient forces can be computed for particles of arbitrary size using MFS (\figref{fig:Sizes}).
In particular, we note that gradient forces dominate for $ka \ll 1$, while the other forces become comparable in magnitude only for $ka \gtrsim 1$.
Although the optical binding force between a pair of particles is $\sim$1 order of magnitude weaker than the scattering force (and has approximately the same scaling with size), this is mitigated by two factors: 1) for clusters of many particles the binding forces will be additive, while the scattering force is not, and 2) the scattering forces can be cancelled with appropriate field design.

In the results that follow, particles are suspended in a pair of counter-propagating acoustic plane waves travelling in the $\pm z$-direction.
This is consistent with the design of common acoustic levitation experiments \cite{wangSoundmediatedStableConfigurations2017a, abdelazizDynamicsAcousticallyTrapped2021, bareschObservationSingleBeamGradient2016}, apart from the fact that we do not assume the beam is strongly focused in the transverse direction.
The interference between these beams confines the particles to a single plane in $z$, but allows them to move freely within this plane without experiencing gradient or scattering forces.
Unless otherwise specified, particles are simulated at a pressure node of the incident field, which has field amplitude of 200 Pa and frequency 40 kHz.

We find it useful to introduce here a reference force, to which one can scale all other relevant effects.
Here, we use the force experienced by a perfectly absorbing sphere within an external acoustic field, $F_0$:
\begin{align}
    F_0 = \frac{p_{1}^{2}}{2\rho_{0}c_{0}^{2}} \pi a^{2},
    \label{eq:RefForce}
\end{align}
where $p_{1}$ is the amplitude of the oscillating external pressure field.

\subsection{Pairwise Interactions}

We first consider the force between pairs of particles in our acoustic standing wave (\figref{fig:Maps}).
In general, we observe an oscillating force -- with a period given by the sound wavelength -- which falls off like $1/r$ and is primarily in the radial direction.
This arises because the main contribution to the acoustic binding force is a gradient force in the combined incoming field and the field scattered from the neighboring particle, whose amplitude falls off like $1/r$.
(Although the binding force can in part be explained as a `gradient force', we do not treat it as part of the gradient force because it is only present for multiple particles.)
Note that the force is generally quite weak if the particles are exactly on the pressure node ($z=0$); this is because the incoming field has no amplitude here.
In practice, gravity will displace particles from the anti-node, in which case this transverse force will be stronger.

For larger particles ($ka \gtrsim 2$), the scattering becomes more intense along the $z$-axis, and so the binding force does as well.
Interestingly, the qualitative features of the acoustic binding force are quite similar to optical binding (\figref{fig:Maps}\textbf{d-f}).
This is explained by the fact that both are second order radiation scattering forces which arise for the same general reasons; the differences can be attributed to the fact that optical radiation is vector while acoustic fields are scalar.
We note also that the optical forces shown in \figref{fig:Maps} are for a relative refractive index of 1.5 -- approximately equivalent to glass particles in air -- while the acoustic forces are computed for an infinite impedance contrast.

The features of either acoustic or optical binding have many characteristics which make them interesting candidates for self- or directed-assembly of particle clusters.
First, we note that the oscillatory nature of the force means that the particles have multiple stable separations whose distance can be tuned by changing the field wavelength.
We characterize this with an effective spring constant, $k_{eff} = -\partial F_r / \partial r$, at each of the stable particle separations (\figref{fig:keff}).
For closely spaced particles, there are additional near-field effects which are strongly dependent on size parameter, $ka$.
Indeed, the binding force can be either attractive or repulsive for contacting spheres (\figref{fig:keff}\textbf{c}), and as a result this is tunable via sound frequency.
Although not considered here, we also note that a sound wave composed of several different frequencies could superimpose the interference patterns and produce an even more complicated -- and tunable -- interparticle force.

A pair of particles in a viscous fluid is effectively a damped harmonic oscillator, and so can be characterized by a dimensionless quality factor, $Q$.
If we assume a Stokes drag law for the particles, we can compute the quality factor as a function of particle size and density (see Supplemental Materials section C).
For experimental parameters relevant for solid particles in a gaseous medium, this quality factor scales like $Q \sim 6\ (ka)^{7/2} \left(\frac{\rho_p }{1 \textrm{kg/m}^3}\right)^{1/2}$ for $ka \lesssim 3$, and has a somewhat more complicated structure for $ka \gtrsim 3$.
(Note that this scaling assumes the incoming wave pressure is increased relative to the gravitational force so that $F_0 = 4.25 mg$.)
As a result, we expect oscillations of particle pairs (or many particle clusters) to be under-damped for realistic experimental parameters, in contrast to optically bound systems which -- assuming wavelength sized particles and visible optical fields -- should be in the over damped regime unless the particles are suspended in vacuum.
As we shall see in section `Evolution of Many Particle Clusters', this has significant implications for the formation of acoustic clusters with many particles.
We also note that a dependence of cluster stability on damping level has previously been predicted for optically bound systems \cite{ngPhotonicClustersFormed2005}.

\subsection{Acoustic Binding of Many Particles}

How does the acoustic binding force scale when we have more than two particles in the field?
The total force on each particle is determined by a combination of the incoming and scattered fields, which manifest in a net force through the $p_2$ term.
As $p_2$ relies on the square of the field values, there are three terms involved in a force calculation: $\phi_{in}^{2}$, $\phi_{sc}^{2}$, and $\phi_{in}\phi_{sc}$.
For weakly scattering particles, or for bodies interacting in the (scattered) far-field one can neglect $\phi_{sc}^{2}$ as it is very small in comparison to the other terms. 
In this case, the force is (approximately) linear in $\phi_{sc}$, in which case one would expect the force to be nearly pairwise; indeed this simplifying assumption has been made in some previous research \cite{zhangAcousticallyMediatedLongrange2016}.
However, for strongly scattering particles which are close together the scattered field can become quite strong ($\phi_{sc} \sim \phi_{inc}$), and so this approximation should break down.

Thus, in the $ka \gtrsim 1$ regime, we can no longer assume that the stable particle separation distances for many particle clusters will be the same as for pairs.
To explore this effect, we compute an effective spring constant for a hexagonal cluster of 7 particles, using the radial force on one of the outer particles (\figref{fig:keff}\textbf{b, d}).
For small particles ($ka \leq 1.8$) we observe that the stable separations shift from $r \sim 1 \lambda$ to $r \sim 1.18 \lambda$.
For larger particles, we observe a more complicated shift in the the stable positions, consistent with the observation that multiple-scattering effects should be stronger for larger particles.
From these observations, we can conclude that a pairwise force approximation is not appropriate for closely spaced wavelength sized particles, and that the full $N$-body force calculations must be performed for accurate results.

\begin{figure}
\begin{centering}
\includegraphics[width = 0.95\linewidth]{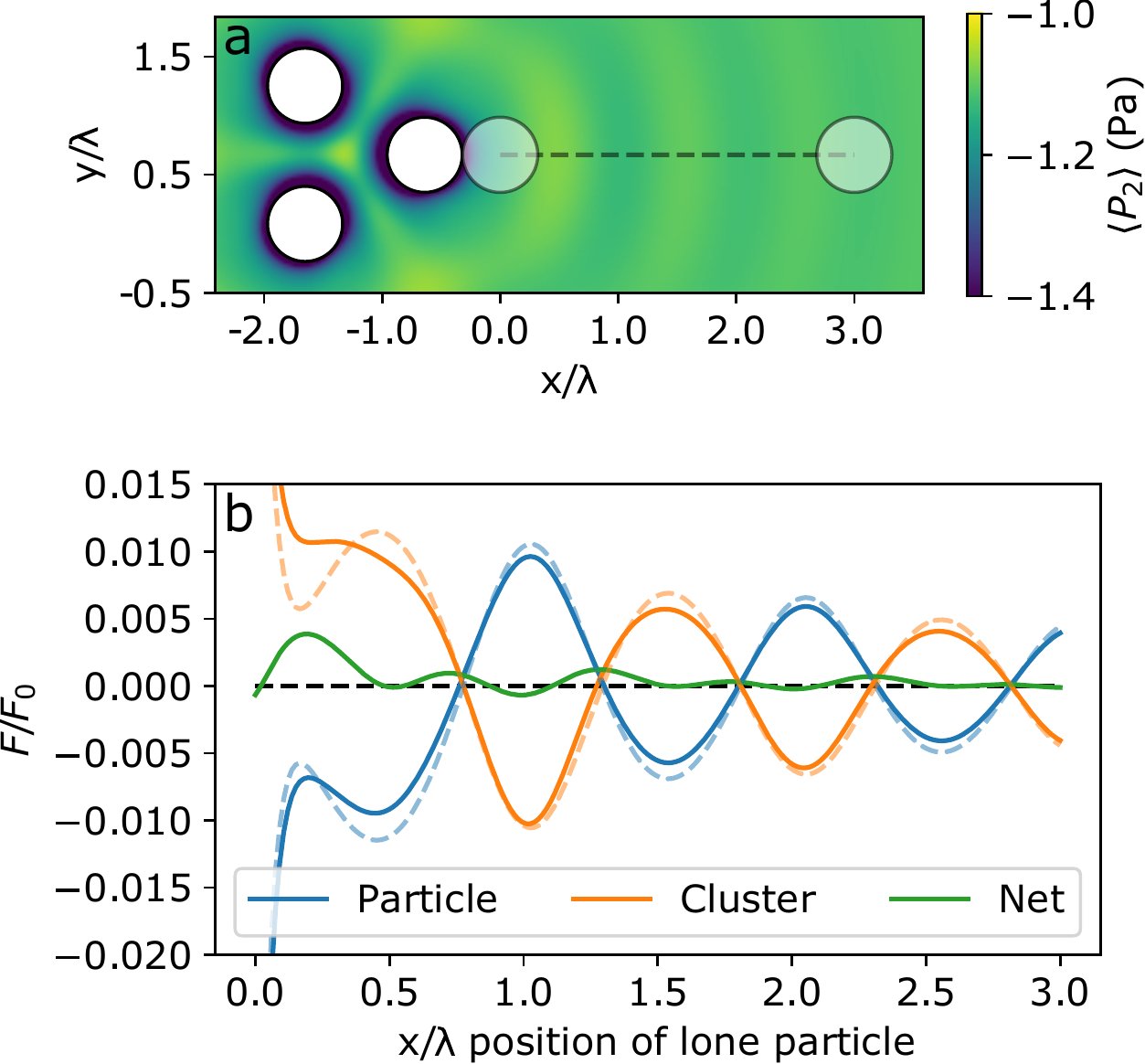}
\end{centering}
\caption{Non-conservative effects in acoustically bound systems. {\bf a)} The second order pressure field, $p_2$, near a stable triangular cluster with $ka=1.466$.
{\bf b)} The force on an extra particle placed near the cluster, as well as the additional force on the cluster itself.
The non-zero net force implies the presence of non-conservative effects which can give rise to drifting of stable clusters in the right configuration. The forces computed from a pairwise model are plotted as dashed lines and show clear discrepancy compared to the actual forces, especially when the lone particle is in close proximity to the cluster. }
\label{fig:Pairwise}
\end{figure}

A pair of particles at the same $z$ coordinate -- but displaced in $x$/$y$ -- must have the equal and opposite forces by symmetry.
For more than two particles, this is no longer the case, especially once multiple scattering effects are included.
To probe for these effects, we consider the forces between a stable three-particle triangular cluster and a fourth particle placed nearby (\figref{fig:Pairwise}\textbf{a}).
If we compare the net forces on the cluster and the lone particle, we find they are not equal and opposite, and this effect becomes stronger as the lone particle gets closer to the cluster. (\figref{fig:Pairwise}\textbf{b}).
Conversely, if we compute the forces by summing over pairs of particles, we find they are equal and opposite, as expected.
Although this non-conservative force appears at first to violate Newton's third law, this is not the case: the net momentum is carried away by the scattered acoustic field.
This simple result demonstrates that acoustic binding can produce non-conservative driving forces, but only when the full $N$-body force is considered.
As we shall show in the next section, it is also possible to produce stable clusters of particles with a non-zero force on the entire cluster.

\subsection{Evolution of Many Particle Clusters}
\begin{table}
\begin{tabular}{| p{3.5cm} p{1.2cm} p{3.5cm}|}
    \multicolumn{3}{c}{Model Parameters} \\
    \hline
    Name & Symbol & Value  \\
    \hline
    Density (Fluid) & $\rho_0$ & 1.225 kg/m$^3$ \\
    Viscosity (Fluid) & $\mu$ & 1.81 $\times$ 10$^{-5}$ kg/(m $\cdot$ s) \\
    Sound Frequency & $f$ & 40 kHz \\
    Sound Amplitude & $p_a$ & *63.25, 200, 632.5 Pa \\
    Sound Speed (Fluid) & $c_0$ & 343 m/s \\
    Sound Speed (Particle) & $c_p$ & *2000, 2500, 3000 m/s \\
    Density (Particle) & $\rho_p$ & *1, 10, 100 kg/m$^3$ \\
    Particle Radius & $a$ & 0.002 mm \\
    Contact Ratio & $\alpha$ & 1.025 \\
    Contact Exponent & $\beta$ & 4 \\
    Contact Prefactor & $C$ & 5 \\
    \hline 
    
\end{tabular}
\caption{Parameters used in the molecular dynamics simulations. Values marked with a star are used respectively within the three sets of 100 trials performed.}
\label{MD Pars}
\end{table}

\begin{figure*}
    \centering
    \includegraphics[width = 0.95\linewidth]{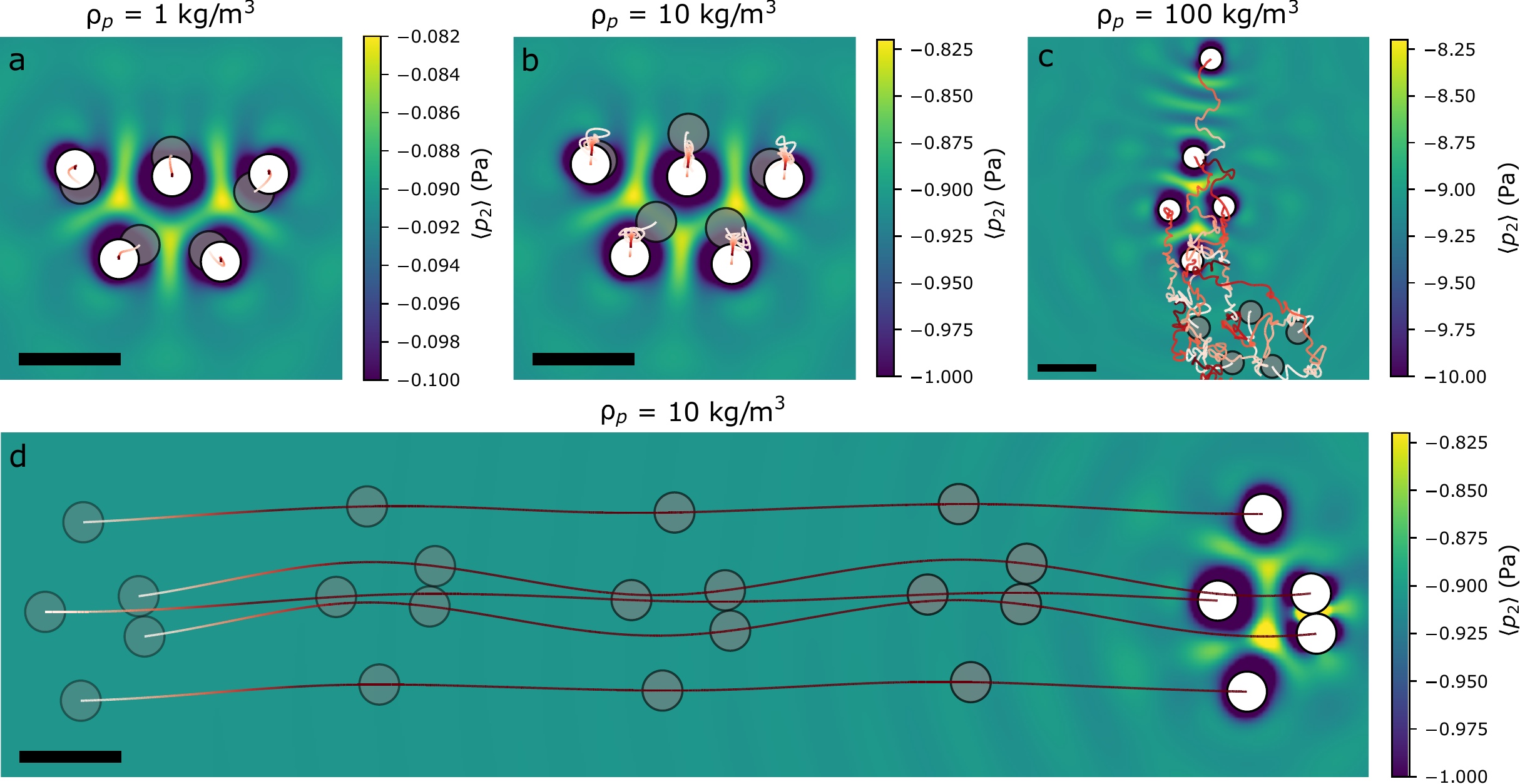}
    \caption{Acoustically bound cluster evolution.
    Initial (gray transparent) and final (white) states produced using simulation data for {\bf a)} highly damped, {\bf b)} mid-damped, and {\bf c)} low-damped systems, all beginning from the same state. 
    The high and mid-damped trials both resulted in the common `boat' configuration, while for lower levels of damping one of the particles is ejected from the cluster.
    The colormap indicates the second order pressure field on the x-y plane surrounding the particles, and the particle tracks are colored sequentially from white to dark red to represent time progression. 
    {\bf d)} A simulation of a quasi-stable `manta-ray' configuration, showing the instantaneous states at various progressive times. Scale bar is 1cm in all panels.
    See supplementary movies 1 and 2 to see `boat' and `manta ray' formations respectively.}
    \label{fig:SimShots}
\end{figure*}

The results described up to this point have been computed only for fixed particle locations.
To understand how many particles freely moving in an acoustic binding force evolve in time, we have coupled our force model to a simple molecular dynamics simulation.
To prevent particle overlap (which would cause the MFS model to produce non-physical results), we also include a short range repulsive force of the form:
\begin{align}
    \V F_{r} &= C F_{0} \left[ \frac{\alpha - \frac{R_{ij}}{2a}}{\alpha-1} \right]^{\beta} \Vh{R}_{ij}
    \label{eq:contact}
\end{align}
This force is designed to turn on slightly before the particles are in actual contact, but to do so at such a distance that it does not significantly affect the results (see Supplemental Materials section C).
We also include a gravitational force ($\V F_g = -9.8 \textrm{ m/s}^2\ \hat z$), and a Stokes drag ($\V F_d = 6 \pi \mu a \V v$), where we assume the background medium is air at STP.
Typical particle velocities are of order 30 mm/s or less, and so the Reynolds number of the flow around a particle is $Re \lesssim 7$; as a result a Stokes drag law is a reasonably good approximation.
We do not consider hydrodynamic coupling between the particles.
Time stepping in this model is handled using an adaptive timestep Runge-Kutta integrator (the Dormand-Prince method), specified to have an absolute velocity error tolerance of $10^{-6}\textrm{m/s}$.
No thermal fluctuations are considered, as they should have negligible impact at these scales.

\begin{figure}
\begin{centering}
\includegraphics[width = 0.9\linewidth]{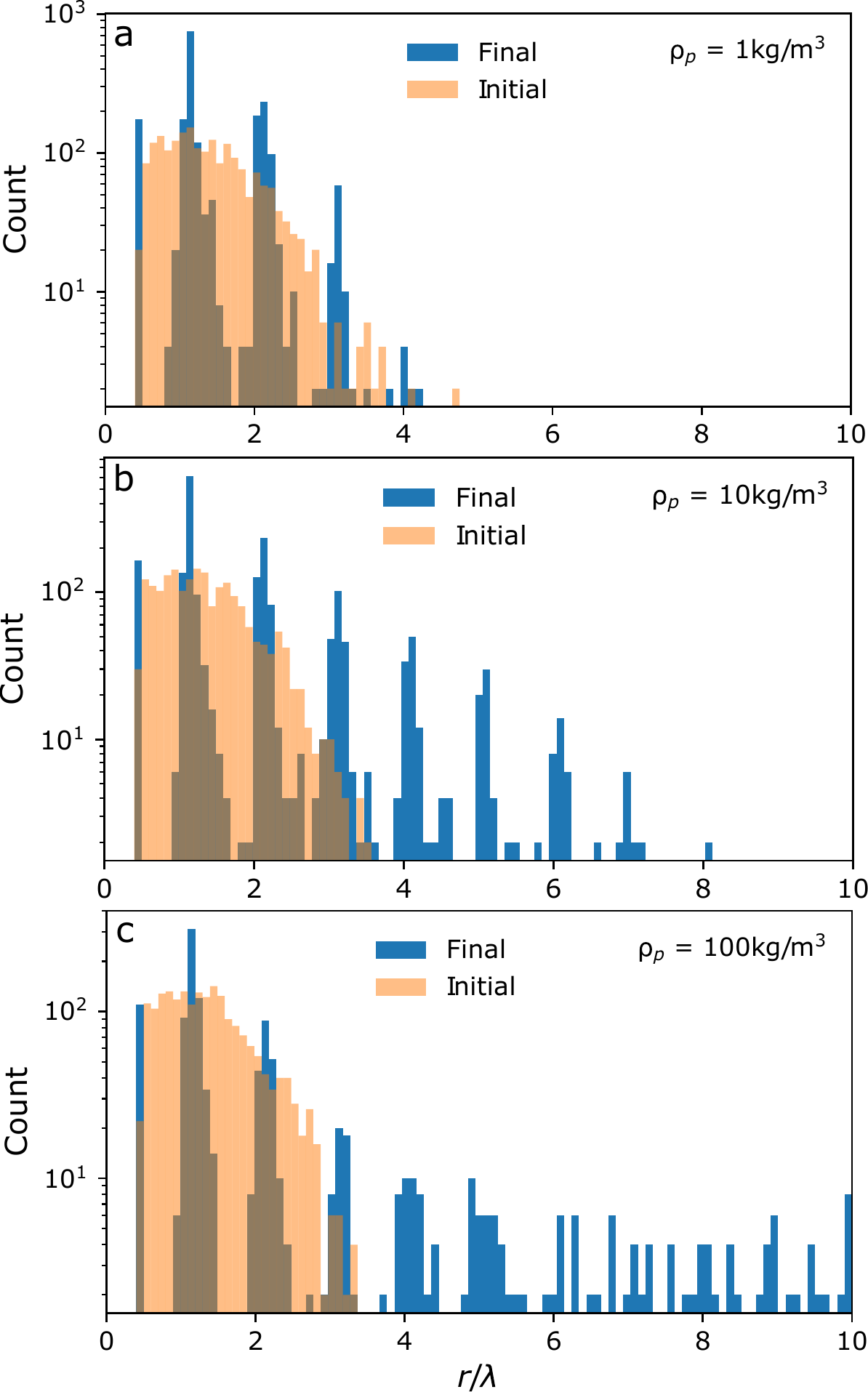}
\end{centering}
\caption{Final interparticle separations. Histograms of the final interparticle separation distances for three sets of 100 trial MD simulations. \textbf{(a)} high damping, \textbf{(b)} mid damping,\textbf{(c)} low damping. In all cases the first preferred separation distance is 1.14 $\lambda$, indicating the importance of multibody effects.
\label{fig:Separation}}
\end{figure}

\begin{figure*}
\begin{centering}
\includegraphics[width = 0.95\linewidth]{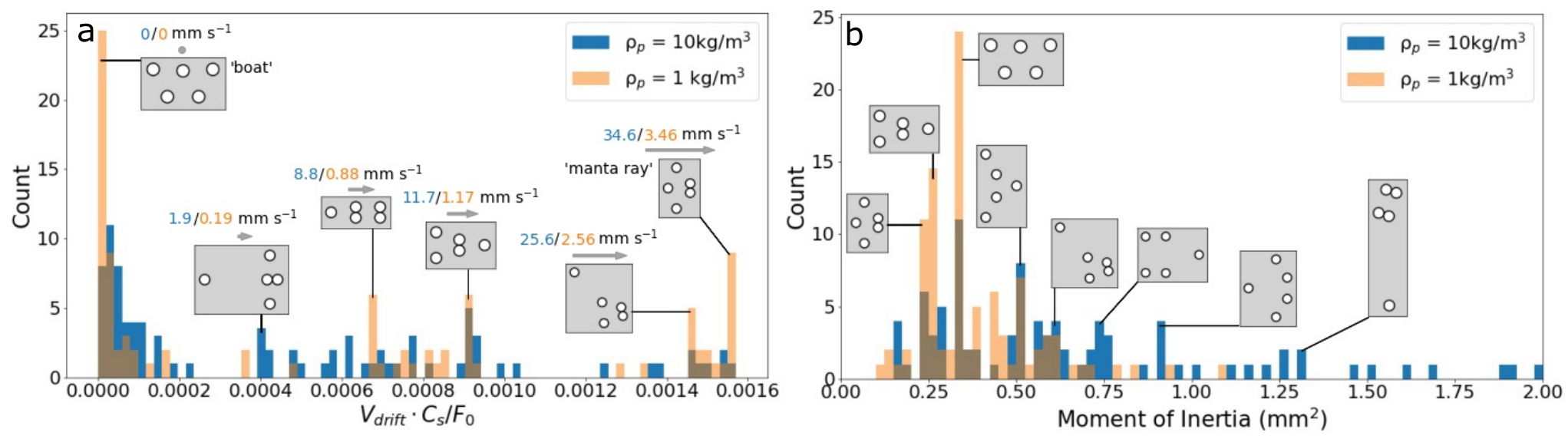}
\end{centering}
\caption{Cluster drift speeds and moments of inertia. Histograms of the drift speed ({\bf a}) and moment of inertia ({\bf b}) of some of the common clusters found in the trial simulations. Blue correspond to mid damping, while orange figures correspond to high damping. Data from trials involving low damping were omitted as the damping was not sufficient to lock particles into a local minimum, i.e. configurations tended to fly apart. \label{fig:MomV}
}
\end{figure*}

The simulations were performed for sets of 100 trials at various damping parameters, with initial particle locations assigned according to a Gaussian distribution with $\sigma = 3a$ (see supplementary materials section B).
All sets of trials were performed for $ka=1.466$, which represents particles of size $a=2$ mm and a driving frequency of 40 kHz in air.
The particle density and incident field amplitude were both varied between trial sets in order to achieve identical binding strengths with varying levels of damping; the density and amplitude values used can be found in Table \ref{MD Pars}
and correspond roughly to Aerogel, expanded polystyrene, and urethane foam, as well as the driving pressures necessary to levitate bodies of these densities.
(As before, the pressure is modulated so that $F_0 = 4.25  mg$ irrespective of density.)
The quality factors for pairs of particles with these properties is $Q \sim $ 11, 34, 107.
In these time-evolution studies we allow for motion in the z-direction so that the gravitational force pulls particles to a natural levitation plane which sits slightly below a pressure node.
The simulations run for a predetermined time (10 s) or until a terminal event is reached, which here corresponds to acceleration of all particles falling below a minimum threshold of $|\V{a}|\leq 10^{-6}$ m/s$^2$.

In general, we observe that randomly placed configurations of 5 particles will arrange themselves into stable clusters, with a strong preference for near-integer wavelength particle spacing (\figref{fig:Separation}).
Interestingly the first preferred interparticle separation distance within all sets of trials converges to $1.14\lambda$, which is slightly greater than the value which would be expected from pairwise interactions.
This separation can be reproduced by instead considering the forces on a 7-particle hexagonal cluster, which has a nearly identical equilibrium separation (\figref{fig:keff}{\bf b}).

Damping plays a critical role in the formation of the clusters, as seen in \figref{fig:SimShots}{\bf a-c}.
As expected, the relaxation time is much shorter for the trials with higher damping.
Although the level of damping should not change the equilibrium configurations, weakly damped systems have difficulty shedding their kinetic energy so they can settle into a stable cluster.
On average, this produces clusters with larger particle separations as the damping is reduced, or, equivalently, the density is increased.
The low-damped trials produced virtually no compact five-particle configurations; at least one particle was generally accelerated through a local minima without enough damping to slow it down, and was effectively ejected away from the rest of the particles in a cluster (e.g.~\figref{fig:SimShots}{\bf c} and supplementary movie 4).
This behavior seems to become prominent when the pairwise quality factor is $Q \gtrsim 50$, and is consistent with results previously seen in simulations of optically bound clusters  \cite{ngPhotonicClustersFormed2005}, which in some cases require a minimum level of damping to form stable structures \cite{liNonHermitianPhysicsOptical2021}.

For all levels of damping, we find that most of the clusters are not stationary, but rather have some non-zero drift velocity when the cluster is asymmetric.
Thus, we characterize the final structures in terms of the moment of inertia and drift speed of the final state of the simulation (\figref{fig:MomV}), with cluster moment of inertia given by:
\begin{align}
    I &= \sum_i^{N_p} (x_i - \bar{x})^2 + (y_i - \bar{y})^2 \label{eq:MomI}
\end{align}
where the sum is over all particles and $\bar{x}/\bar{y}$ denote the center of mass of the configuration in the x/y direction.
A number of cluster geometries were commonly produced throughout the mid and high damped trials, with the most common shape being what we term the `boat' configuration (\figref{fig:SimShots}{\bf a,b}).
The boat -- which is one of the few observed clusters with no drift velocity -- comprised 12\% of the mid-damped and 18\% of the high damped final configurations (videos of this and other structures forming can be seen in the supplemental materials).
We do not observe a correlation between moment of inertia and drift speed, suggesting that extended particle clusters are not required to produce drifting configurations.

Additionally, the drift speed can be rescaled by a reference velocity which would be experienced by a single particle subject to a force of $F_0$ -- this is equivalent to the speed a perfectly sound absorbing particle would move in a single plane wave of sound.
Notably, the fastest moving clusters all contain a pair of particles in contact, which may effectively behave like a larger particle and so experience higher non-conservative driving forces.

In rarer cases we also observe quasi-stable clusters which oscillate or rotate in time; an example is shown in \figref{fig:SimShots}{\bf d} and supplementary movie 2.
In this configuration -- which we term the `manta ray' -- the cluster itself undergoes oscillations while it drifts in space.
This oscillatory motion continues indefinitely with no perceptual loss in amplitude, with the driving of the local oscillations perfectly balanced by the system damping.
The oscillatory behavior of the manta ray configuration is only observed only for the mid-damped case; the high-damped case produces nearly identical configurations: these do not oscillate but do drift at a high velocity.

\section{Discussion}

Acoustic binding produces complex, long range forces which make it an intriguing candidate for directed assembly of mm-scale systems.
Although previous work has considered simplified models of this force \cite{zhangAcousticallyMediatedLongrange2016,garcia-sabateExperimentalStudyInterparticle2014,baaschMultibodyDynamicsAcoustophoresis2017}, we find that non-pairwise and non-conservative effects significantly modify the resulting structures and their stability.
This is especially true in the size regime where the multi-particle binding force is strongest compared to gradient and scattering forces, $ka \sim 1-10$.
If one wishes to use this force for on-demand assembly of complex structures, this has both positive and negative attributes.
On one hand, the modifications of the force due to multiple-scattering effects could be exploited to assemble more complex structures without using a more complex incoming field.
Moreover, the non-conservative driving forces make acoustically bound clusters analogous to active-matter systems which are currently being explored for a variety of applications \cite{kangRecentProgressActive2019, sanchezSpontaneousMotionHierarchically2012, ramaswamyActiveMatter2017}.
Unlike typical active matter systems, these driving forces are the result of specific particle configurations, and so -- in principle -- they are both switchable and tunable.
It is likely that non-spherical particles or even mixtures of differently sized particles could enhance these effects, as they arise due to symmetry breaking of the stable configurations.

Conversely, the observation that the many-body forces must be considered raises practical difficulties in computing the acoustic binding forces and predicting the structures they produce.
This will be especially true as the number of particles is increased, since the computational cost of a fully coupled $N$-body system scales like $N^3$.
Although this could be somewhat mitigated using advanced numerical techniques (such as the fast multiple method \cite{darveFastMultipoleMethod2000a} or GPU-based simulations), it may also be possible to produce empirical models of the many-body force which are more computationally efficient.
In either case, with further numerical optimization these models could allow acoustic binding to be used for `inverse-design' approaches where an incoming field (or combination of several) is used to produce a desired structure on demand \cite{kumarInverseDesignCharged2019, adorfInverseDesignSimple2018, lindquistCommunicationInverseDesign2016b}.

Finally, we note that -- to our knowledge -- non-contact acoustic binding effects have only been observed in a single experiment, and the largest clusters observed had only 3 particles \cite{wangSoundmediatedStableConfigurations2017a}.
There are likely two reasons for this:
1) most acoustic levitation experiments used focused fields which do not have room for particles that  are transversely displaced by one or more wavelengths, and 
2) driving forces make configurations of solid particles ($\rho \sim 1000$ kg/m$^3$) unstable in air.
Both of these limitations could be overcome with appropriate experimental design.
Indeed, there is no fundamental requirement to use focused sound fields for acoustic levitation.
Moreover, the relative amount of damping can be modified either by using a different background medium (e.g.~a liquid rather than a gas, as used in the aforementioned experiment), or using higher frequency sound waves and smaller particles.
Such a system would have both fundamental and practical applications.
Understanding the self-organization of actively driven systems has attracted considerable interest in the last decade; acoustic binding provides a tunable platform to study these effects on easily accessible length scales.
With continued research, acoustic binding could ultimately provide a practical method for controlling the assembly of particle clusters or even meta-materials composed of large numbers of particles in active or passive configurations.

\vspace{1cm}

\bibliography{MFS.bib}

\begin{thebibliography}{62}%
\makeatletter
\providecommand \@ifxundefined [1]{%
 \@ifx{#1\undefined}
}%
\providecommand \@ifnum [1]{%
 \ifnum #1\expandafter \@firstoftwo
 \else \expandafter \@secondoftwo
 \fi
}%
\providecommand \@ifx [1]{%
 \ifx #1\expandafter \@firstoftwo
 \else \expandafter \@secondoftwo
 \fi
}%
\providecommand \natexlab [1]{#1}%
\providecommand \enquote  [1]{``#1''}%
\providecommand \bibnamefont  [1]{#1}%
\providecommand \bibfnamefont [1]{#1}%
\providecommand \citenamefont [1]{#1}%
\providecommand \href@noop [0]{\@secondoftwo}%
\providecommand \href [0]{\begingroup \@sanitize@url \@href}%
\providecommand \@href[1]{\@@startlink{#1}\@@href}%
\providecommand \@@href[1]{\endgroup#1\@@endlink}%
\providecommand \@sanitize@url [0]{\catcode `\\12\catcode `\$12\catcode
  `\&12\catcode `\#12\catcode `\^12\catcode `\_12\catcode `\%12\relax}%
\providecommand \@@startlink[1]{}%
\providecommand \@@endlink[0]{}%
\providecommand \url  [0]{\begingroup\@sanitize@url \@url }%
\providecommand \@url [1]{\endgroup\@href {#1}{\urlprefix }}%
\providecommand \urlprefix  [0]{URL }%
\providecommand \Eprint [0]{\href }%
\providecommand \doibase [0]{https://doi.org/}%
\providecommand \selectlanguage [0]{\@gobble}%
\providecommand \bibinfo  [0]{\@secondoftwo}%
\providecommand \bibfield  [0]{\@secondoftwo}%
\providecommand \translation [1]{[#1]}%
\providecommand \BibitemOpen [0]{}%
\providecommand \bibitemStop [0]{}%
\providecommand \bibitemNoStop [0]{.\EOS\space}%
\providecommand \EOS [0]{\spacefactor3000\relax}%
\providecommand \BibitemShut  [1]{\csname bibitem#1\endcsname}%
\let\auto@bib@innerbib\@empty
\bibitem [{\citenamefont {Perry}\ \emph {et~al.}(2015)\citenamefont {Perry},
  \citenamefont {{Holmes-Cerfon}}, \citenamefont {Brenner},\ and\ \citenamefont
  {Manoharan}}]{perryTwoDimensionalClustersColloidal2015}%
  \BibitemOpen
  \bibfield  {author} {\bibinfo {author} {\bibfnamefont {R.~W.}\ \bibnamefont
  {Perry}}, \bibinfo {author} {\bibfnamefont {M.~C.}\ \bibnamefont
  {{Holmes-Cerfon}}}, \bibinfo {author} {\bibfnamefont {M.~P.}\ \bibnamefont
  {Brenner}},\ and\ \bibinfo {author} {\bibfnamefont {V.~N.}\ \bibnamefont
  {Manoharan}},\ }\bibfield  {title} {\bibinfo {title} {Two-{{Dimensional
  Clusters}} of {{Colloidal Spheres}}: Ground {{States}}, {{Excited States}},
  and {{Structural Rearrangements}}},\ }\href@noop {} {\bibfield  {journal}
  {\bibinfo  {journal} {Phys. Rev. Lett.}\ }\textbf {\bibinfo {volume} {114}}
  (\bibinfo {year} {2015})}\BibitemShut {NoStop}%
\bibitem [{\citenamefont {Meng}\ \emph {et~al.}(2010)\citenamefont {Meng},
  \citenamefont {Arkus}, \citenamefont {Brenner},\ and\ \citenamefont
  {Manoharan}}]{mengFreeEnergyLandscapeClusters2010}%
  \BibitemOpen
  \bibfield  {author} {\bibinfo {author} {\bibfnamefont {G.}~\bibnamefont
  {Meng}}, \bibinfo {author} {\bibfnamefont {N.}~\bibnamefont {Arkus}},
  \bibinfo {author} {\bibfnamefont {M.~P.}\ \bibnamefont {Brenner}},\ and\
  \bibinfo {author} {\bibfnamefont {V.~N.}\ \bibnamefont {Manoharan}},\
  }\bibfield  {title} {\bibinfo {title} {The {{Free}}-{{Energy Landscape}} of
  {{Clusters}} of {{Attractive Hard Spheres}}},\ }\href@noop {} {\bibfield
  {journal} {\bibinfo  {journal} {Sci. Mag.}\ } (\bibinfo {year}
  {2010})}\BibitemShut {NoStop}%
\bibitem [{\citenamefont {Lenshof}\ \emph {et~al.}(2012)\citenamefont
  {Lenshof}, \citenamefont {Magnusson},\ and\ \citenamefont
  {Laurell}}]{lenshofAcoustofluidicsApplicationsAcoustophoresis2012}%
  \BibitemOpen
  \bibfield  {author} {\bibinfo {author} {\bibfnamefont {A.}~\bibnamefont
  {Lenshof}}, \bibinfo {author} {\bibfnamefont {C.}~\bibnamefont {Magnusson}},\
  and\ \bibinfo {author} {\bibfnamefont {T.}~\bibnamefont {Laurell}},\
  }\bibfield  {title} {\bibinfo {title} {Acoustofluidics 8: Applications of
  acoustophoresis in continuous flow microsystems},\ }\href
  {https://doi.org/10.1039/c2lc21256k} {\bibfield  {journal} {\bibinfo
  {journal} {Lab Chip}\ }\textbf {\bibinfo {volume} {12}},\ \bibinfo {pages}
  {1210} (\bibinfo {year} {2012})}\BibitemShut {NoStop}%
\bibitem [{\citenamefont {Neuman}\ and\ \citenamefont
  {Block}(2004)}]{neumanOpticalTrapping2004}%
  \BibitemOpen
  \bibfield  {author} {\bibinfo {author} {\bibfnamefont {K.~C.}\ \bibnamefont
  {Neuman}}\ and\ \bibinfo {author} {\bibfnamefont {S.~M.}\ \bibnamefont
  {Block}},\ }\bibfield  {title} {\bibinfo {title} {Optical trapping},\ }\href
  {https://doi.org/10.1063/1.1785844} {\bibfield  {journal} {\bibinfo
  {journal} {Review of Scientific Instruments}\ }\textbf {\bibinfo {volume}
  {75}},\ \bibinfo {pages} {2787} (\bibinfo {year} {2004})}\BibitemShut
  {NoStop}%
\bibitem [{\citenamefont
  {Ashkin}(1970)}]{ashkinAccelerationTrappingParticles1970}%
  \BibitemOpen
  \bibfield  {author} {\bibinfo {author} {\bibfnamefont {A.}~\bibnamefont
  {Ashkin}},\ }\bibfield  {title} {\bibinfo {title} {Acceleration and
  {{Trapping}} of {{Particles}} by {{Radiation Pressure}}},\ }\href@noop {}
  {\bibfield  {journal} {\bibinfo  {journal} {Phys. Rev. Lett.}\ }\textbf
  {\bibinfo {volume} {24}},\ \bibinfo {pages} {156} (\bibinfo {year}
  {1970})}\BibitemShut {NoStop}%
\bibitem [{\citenamefont {Marag{\`o}}\ \emph {et~al.}(2013)\citenamefont
  {Marag{\`o}}, \citenamefont {Jones}, \citenamefont {Gucciardi}, \citenamefont
  {Volpe},\ and\ \citenamefont
  {Ferrari}}]{maragoOpticalTrappingManipulation2013}%
  \BibitemOpen
  \bibfield  {author} {\bibinfo {author} {\bibfnamefont {O.~M.}\ \bibnamefont
  {Marag{\`o}}}, \bibinfo {author} {\bibfnamefont {P.~H.}\ \bibnamefont
  {Jones}}, \bibinfo {author} {\bibfnamefont {P.~G.}\ \bibnamefont
  {Gucciardi}}, \bibinfo {author} {\bibfnamefont {G.}~\bibnamefont {Volpe}},\
  and\ \bibinfo {author} {\bibfnamefont {A.~C.}\ \bibnamefont {Ferrari}},\
  }\bibfield  {title} {\bibinfo {title} {Optical trapping and manipulation of
  nanostructures},\ }\href {https://doi.org/10.1038/nnano.2013.208} {\bibfield
  {journal} {\bibinfo  {journal} {Nature Nanotech}\ }\textbf {\bibinfo {volume}
  {8}},\ \bibinfo {pages} {807} (\bibinfo {year} {2013})}\BibitemShut {NoStop}%
\bibitem [{\citenamefont {Ashkin}\ \emph {et~al.}(1987)\citenamefont {Ashkin},
  \citenamefont {Dziedzic},\ and\ \citenamefont
  {Yamane}}]{ashkinOpticalTrappingManipulation1987}%
  \BibitemOpen
  \bibfield  {author} {\bibinfo {author} {\bibfnamefont {A.}~\bibnamefont
  {Ashkin}}, \bibinfo {author} {\bibfnamefont {J.~M.}\ \bibnamefont
  {Dziedzic}},\ and\ \bibinfo {author} {\bibfnamefont {T.}~\bibnamefont
  {Yamane}},\ }\bibfield  {title} {\bibinfo {title} {Optical trapping and
  manipulation of single cells using infrared laser beams},\ }\href
  {https://doi.org/10.1038/330769a0} {\bibfield  {journal} {\bibinfo  {journal}
  {Nature}\ }\textbf {\bibinfo {volume} {330}},\ \bibinfo {pages} {769}
  (\bibinfo {year} {1987})}\BibitemShut {NoStop}%
\bibitem [{\citenamefont {Ashok}\ and\ \citenamefont
  {Dholakia}(2012)}]{ashokOpticalTrappingAnalytical2012}%
  \BibitemOpen
  \bibfield  {author} {\bibinfo {author} {\bibfnamefont {P.~C.}\ \bibnamefont
  {Ashok}}\ and\ \bibinfo {author} {\bibfnamefont {K.}~\bibnamefont
  {Dholakia}},\ }\bibfield  {title} {\bibinfo {title} {Optical trapping for
  analytical biotechnology},\ }\href
  {https://doi.org/10.1016/j.copbio.2011.11.011} {\bibfield  {journal}
  {\bibinfo  {journal} {Current Opinion in Biotechnology}\ }\textbf {\bibinfo
  {volume} {23}},\ \bibinfo {pages} {16} (\bibinfo {year} {2012})}\BibitemShut
  {NoStop}%
\bibitem [{\citenamefont {Pang}\ and\ \citenamefont
  {Gordon}(2012)}]{pangOpticalTrappingSingle2012}%
  \BibitemOpen
  \bibfield  {author} {\bibinfo {author} {\bibfnamefont {Y.}~\bibnamefont
  {Pang}}\ and\ \bibinfo {author} {\bibfnamefont {R.}~\bibnamefont {Gordon}},\
  }\bibfield  {title} {\bibinfo {title} {Optical {{Trapping}} of a {{Single
  Protein}}},\ }\href {https://doi.org/10.1021/nl203719v} {\bibfield  {journal}
  {\bibinfo  {journal} {Nano Lett.}\ }\textbf {\bibinfo {volume} {12}},\
  \bibinfo {pages} {402} (\bibinfo {year} {2012})}\BibitemShut {NoStop}%
\bibitem [{\citenamefont {Woerdemann}\ \emph {et~al.}(2013)\citenamefont
  {Woerdemann}, \citenamefont {Alpmann}, \citenamefont {Esseling},\ and\
  \citenamefont {Denz}}]{woerdemannAdvancedOpticalTrapping2013}%
  \BibitemOpen
  \bibfield  {author} {\bibinfo {author} {\bibfnamefont {M.}~\bibnamefont
  {Woerdemann}}, \bibinfo {author} {\bibfnamefont {C.}~\bibnamefont {Alpmann}},
  \bibinfo {author} {\bibfnamefont {M.}~\bibnamefont {Esseling}},\ and\
  \bibinfo {author} {\bibfnamefont {C.}~\bibnamefont {Denz}},\ }\bibfield
  {title} {\bibinfo {title} {Advanced optical trapping by complex beam shaping:
  Advanced optical trapping},\ }\href {https://doi.org/10.1002/lpor.201200058}
  {\bibfield  {journal} {\bibinfo  {journal} {Laser \& Photonics Reviews}\
  }\textbf {\bibinfo {volume} {7}},\ \bibinfo {pages} {839} (\bibinfo {year}
  {2013})}\BibitemShut {NoStop}%
\bibitem [{\citenamefont {Burns}\ \emph {et~al.}(1989)\citenamefont {Burns},
  \citenamefont {Fournier},\ and\ \citenamefont
  {Golovchenko}}]{burnsOpticalBinding1989b}%
  \BibitemOpen
  \bibfield  {author} {\bibinfo {author} {\bibfnamefont {M.~M.}\ \bibnamefont
  {Burns}}, \bibinfo {author} {\bibfnamefont {J.-M.}\ \bibnamefont
  {Fournier}},\ and\ \bibinfo {author} {\bibfnamefont {J.~A.}\ \bibnamefont
  {Golovchenko}},\ }\bibfield  {title} {\bibinfo {title} {Optical
  {{Binding}}},\ }\href@noop {} {\bibfield  {journal} {\bibinfo  {journal}
  {Phys. Rev. Lett.}\ }\textbf {\bibinfo {volume} {63}},\ \bibinfo {pages}
  {1234} (\bibinfo {year} {1989})}\BibitemShut {NoStop}%
\bibitem [{\citenamefont {Ng}\ \emph {et~al.}(2005)\citenamefont {Ng},
  \citenamefont {Lin}, \citenamefont {Chan},\ and\ \citenamefont
  {Sheng}}]{ngPhotonicClustersFormed2005}%
  \BibitemOpen
  \bibfield  {author} {\bibinfo {author} {\bibfnamefont {J.}~\bibnamefont
  {Ng}}, \bibinfo {author} {\bibfnamefont {Z.~F.}\ \bibnamefont {Lin}},
  \bibinfo {author} {\bibfnamefont {C.~T.}\ \bibnamefont {Chan}},\ and\
  \bibinfo {author} {\bibfnamefont {P.}~\bibnamefont {Sheng}},\ }\bibfield
  {title} {\bibinfo {title} {Photonic clusters formed by dielectric
  microspheres: Numerical simulations},\ }\href
  {https://doi.org/10.1103/PhysRevB.72.085130} {\bibfield  {journal} {\bibinfo
  {journal} {Physical Review B}\ }\textbf {\bibinfo {volume} {72}},\ \bibinfo
  {pages} {085130} (\bibinfo {year} {2005})}\BibitemShut {NoStop}%
\bibitem [{\citenamefont {Moore}\ and\ \citenamefont
  {Geraci}(2021)}]{mooreSearchingNewPhysics2021}%
  \BibitemOpen
  \bibfield  {author} {\bibinfo {author} {\bibfnamefont {D.~C.}\ \bibnamefont
  {Moore}}\ and\ \bibinfo {author} {\bibfnamefont {A.~A.}\ \bibnamefont
  {Geraci}},\ }\bibfield  {title} {\bibinfo {title} {Searching for new physics
  using optically levitated sensors},\ }\href
  {https://doi.org/10.1088/2058-9565/abcf8a} {\bibfield  {journal} {\bibinfo
  {journal} {Quantum Sci. Technol.}\ }\textbf {\bibinfo {volume} {6}},\
  \bibinfo {pages} {014008} (\bibinfo {year} {2021})}\BibitemShut {NoStop}%
\bibitem [{\citenamefont
  {Bruus}(2012{\natexlab{a}})}]{bruusAcoustofluidicsAcousticRadiation2012b}%
  \BibitemOpen
  \bibfield  {author} {\bibinfo {author} {\bibfnamefont {H.}~\bibnamefont
  {Bruus}},\ }\bibfield  {title} {\bibinfo {title} {Acoustofluidics 7: The
  acoustic radiation force on small particles},\ }\href
  {https://doi.org/10.1039/C2LC21068A} {\bibfield  {journal} {\bibinfo
  {journal} {Lab Chip}\ }\textbf {\bibinfo {volume} {12}},\ \bibinfo {pages}
  {1014} (\bibinfo {year} {2012}{\natexlab{a}})}\BibitemShut {NoStop}%
\bibitem [{\citenamefont {Rabaud}\ \emph {et~al.}(2011)\citenamefont {Rabaud},
  \citenamefont {Thibault}, \citenamefont {Mathieu},\ and\ \citenamefont
  {Marmottant}}]{rabaudAcousticallyBoundMicrofluidic2011}%
  \BibitemOpen
  \bibfield  {author} {\bibinfo {author} {\bibfnamefont {D.}~\bibnamefont
  {Rabaud}}, \bibinfo {author} {\bibfnamefont {P.}~\bibnamefont {Thibault}},
  \bibinfo {author} {\bibfnamefont {M.}~\bibnamefont {Mathieu}},\ and\ \bibinfo
  {author} {\bibfnamefont {P.}~\bibnamefont {Marmottant}},\ }\bibfield  {title}
  {\bibinfo {title} {Acoustically {{Bound Microfluidic Bubble Crystals}}},\
  }\bibfield  {journal} {\bibinfo  {journal} {Phys. Rev. Lett.}\ }\textbf
  {\bibinfo {volume} {106}},\ \href
  {https://doi.org/http://dx.doi.org/10.1103/PhysRevLett.106.134501}
  {http://dx.doi.org/10.1103/PhysRevLett.106.134501} (\bibinfo {year}
  {2011})\BibitemShut {NoStop}%
\bibitem [{\citenamefont {Lim}\ \emph {et~al.}(2019)\citenamefont {Lim},
  \citenamefont {Souslov}, \citenamefont {Vitelli},\ and\ \citenamefont
  {Jaeger}}]{limClusterFormationAcoustic2019}%
  \BibitemOpen
  \bibfield  {author} {\bibinfo {author} {\bibfnamefont {M.~X.}\ \bibnamefont
  {Lim}}, \bibinfo {author} {\bibfnamefont {A.}~\bibnamefont {Souslov}},
  \bibinfo {author} {\bibfnamefont {V.}~\bibnamefont {Vitelli}},\ and\ \bibinfo
  {author} {\bibfnamefont {H.~M.}\ \bibnamefont {Jaeger}},\ }\bibfield  {title}
  {\bibinfo {title} {Cluster formation by acoustic forces and active
  fluctuations in levitated granular matter},\ }\href
  {https://doi.org/10.1038/s41567-019-0440-9} {\bibfield  {journal} {\bibinfo
  {journal} {Nature Physics}\ }\textbf {\bibinfo {volume} {15}},\ \bibinfo
  {pages} {460} (\bibinfo {year} {2019})}\BibitemShut {NoStop}%
\bibitem [{\citenamefont {Jiao}\ \emph {et~al.}(2013)\citenamefont {Jiao},
  \citenamefont {He}, \citenamefont {Leong}, \citenamefont {Kentish},
  \citenamefont {Ashokkumar}, \citenamefont {Manasseh},\ and\ \citenamefont
  {Lee}}]{jiaoExperimentalTheoreticalStudies2013}%
  \BibitemOpen
  \bibfield  {author} {\bibinfo {author} {\bibfnamefont {J.}~\bibnamefont
  {Jiao}}, \bibinfo {author} {\bibfnamefont {Y.}~\bibnamefont {He}}, \bibinfo
  {author} {\bibfnamefont {T.}~\bibnamefont {Leong}}, \bibinfo {author}
  {\bibfnamefont {S.~E.}\ \bibnamefont {Kentish}}, \bibinfo {author}
  {\bibfnamefont {M.}~\bibnamefont {Ashokkumar}}, \bibinfo {author}
  {\bibfnamefont {R.}~\bibnamefont {Manasseh}},\ and\ \bibinfo {author}
  {\bibfnamefont {J.}~\bibnamefont {Lee}},\ }\bibfield  {title} {\bibinfo
  {title} {Experimental and {{Theoretical Studies}} on the {{Movements}} of
  {{Two Bubbles}} in an {{Acoustic Standing Wave Field}}},\ }\href@noop {}
  {\bibfield  {journal} {\bibinfo  {journal} {J. Phys. Chem. B}\ }\textbf
  {\bibinfo {volume} {117}},\ \bibinfo {pages} {12549} (\bibinfo {year}
  {2013})}\BibitemShut {NoStop}%
\bibitem [{\citenamefont {Petersson}\ \emph {et~al.}(2007)\citenamefont
  {Petersson}, \citenamefont {{\AA}berg}, \citenamefont {{Sw{\"a}rd-Nilsson}},\
  and\ \citenamefont {Laurell}}]{peterssonFreeFlowAcoustophoresis2007}%
  \BibitemOpen
  \bibfield  {author} {\bibinfo {author} {\bibfnamefont {F.}~\bibnamefont
  {Petersson}}, \bibinfo {author} {\bibfnamefont {L.}~\bibnamefont
  {{\AA}berg}}, \bibinfo {author} {\bibfnamefont {A.-M.}\ \bibnamefont
  {{Sw{\"a}rd-Nilsson}}},\ and\ \bibinfo {author} {\bibfnamefont
  {T.}~\bibnamefont {Laurell}},\ }\bibfield  {title} {\bibinfo {title} {Free
  {{Flow Acoustophoresis}}: Microfluidic-{{Based Mode}} of {{Particle}} and
  {{Cell Separation}}},\ }\href {https://doi.org/10.1021/ac070444e} {\bibfield
  {journal} {\bibinfo  {journal} {Anal. Chem.}\ }\textbf {\bibinfo {volume}
  {79}},\ \bibinfo {pages} {5117} (\bibinfo {year} {2007})}\BibitemShut
  {NoStop}%
\bibitem [{\citenamefont {Sepehrirahnama}\ \emph {et~al.}(2015)\citenamefont
  {Sepehrirahnama}, \citenamefont {Lim},\ and\ \citenamefont
  {Chau}}]{sepehrirahnamaNumericalStudyInterparticle2015}%
  \BibitemOpen
  \bibfield  {author} {\bibinfo {author} {\bibfnamefont {S.}~\bibnamefont
  {Sepehrirahnama}}, \bibinfo {author} {\bibfnamefont {K.-M.}\ \bibnamefont
  {Lim}},\ and\ \bibinfo {author} {\bibfnamefont {F.~S.}\ \bibnamefont
  {Chau}},\ }\bibfield  {title} {\bibinfo {title} {Numerical study of
  interparticle radiation force acting on rigid spheres in a standing wave},\
  }\href@noop {} {\bibfield  {journal} {\bibinfo  {journal} {ASA}\ }\textbf
  {\bibinfo {volume} {137}} (\bibinfo {year} {2015})}\BibitemShut {NoStop}%
\bibitem [{\citenamefont {Courtney}\ \emph {et~al.}(2013)\citenamefont
  {Courtney}, \citenamefont {Drinkwater}, \citenamefont {Demore}, \citenamefont
  {Cochran}, \citenamefont {Grinenko},\ and\ \citenamefont
  {Wilcox}}]{courtneyDexterousManipulationMicroparticles2013}%
  \BibitemOpen
  \bibfield  {author} {\bibinfo {author} {\bibfnamefont {C.~R.~P.}\
  \bibnamefont {Courtney}}, \bibinfo {author} {\bibfnamefont {B.~W.}\
  \bibnamefont {Drinkwater}}, \bibinfo {author} {\bibfnamefont {C.~E.~M.}\
  \bibnamefont {Demore}}, \bibinfo {author} {\bibfnamefont {S.}~\bibnamefont
  {Cochran}}, \bibinfo {author} {\bibfnamefont {A.}~\bibnamefont {Grinenko}},\
  and\ \bibinfo {author} {\bibfnamefont {P.~D.}\ \bibnamefont {Wilcox}},\
  }\bibfield  {title} {\bibinfo {title} {Dexterous manipulation of
  microparticles using {{Bessel}}-function acoustic pressure fields},\ }\href
  {https://doi.org/10.1063/1.4798584} {\bibfield  {journal} {\bibinfo
  {journal} {Appl. Phys. Lett.}\ }\textbf {\bibinfo {volume} {102}},\ \bibinfo
  {pages} {123508} (\bibinfo {year} {2013})}\BibitemShut {NoStop}%
\bibitem [{\citenamefont {Baasch}\ and\ \citenamefont
  {Dual}(2018)}]{baaschAcoustofluidicParticleDynamics2018}%
  \BibitemOpen
  \bibfield  {author} {\bibinfo {author} {\bibfnamefont {T.}~\bibnamefont
  {Baasch}}\ and\ \bibinfo {author} {\bibfnamefont {J.}~\bibnamefont {Dual}},\
  }\bibfield  {title} {\bibinfo {title} {Acoustofluidic particle dynamics:
  Beyond the {{Rayleigh}} limit},\ }\href {https://doi.org/10.1121/1.5021339}
  {\bibfield  {journal} {\bibinfo  {journal} {The Journal of the Acoustical
  Society of America}\ }\textbf {\bibinfo {volume} {143}},\ \bibinfo {pages}
  {509} (\bibinfo {year} {2018})}\BibitemShut {NoStop}%
\bibitem [{\citenamefont
  {Doinikov}(2001)}]{doinikovAcousticRadiationInterparticle2001}%
  \BibitemOpen
  \bibfield  {author} {\bibinfo {author} {\bibfnamefont {A.}~\bibnamefont
  {Doinikov}},\ }\bibfield  {title} {\bibinfo {title} {Acoustic radiation
  interparticle forces in a compressible fluid},\ }\href
  {https://doi.org/10.1017/S0022112001005055} {\bibfield  {journal} {\bibinfo
  {journal} {J. fluid Mech.}\ }\textbf {\bibinfo {volume} {444}},\ \bibinfo
  {pages} {1} (\bibinfo {year} {2001})}\BibitemShut {NoStop}%
\bibitem [{\citenamefont {Zheng}\ and\ \citenamefont
  {Apfel}(1994)}]{zhengAcousticInteractionForces1994}%
  \BibitemOpen
  \bibfield  {author} {\bibinfo {author} {\bibfnamefont {X.}~\bibnamefont
  {Zheng}}\ and\ \bibinfo {author} {\bibfnamefont {R.~E.}\ \bibnamefont
  {Apfel}},\ }\bibfield  {title} {\bibinfo {title} {Acoustic interaction forces
  between two fluid spheres in an acoustic field},\ }\href@noop {} {\bibfield
  {journal} {\bibinfo  {journal} {ASA}\ }\textbf {\bibinfo {volume} {97}}
  (\bibinfo {year} {1994})}\BibitemShut {NoStop}%
\bibitem [{\citenamefont {Wu}(1991)}]{wuAcousticalTweezers1991}%
  \BibitemOpen
  \bibfield  {author} {\bibinfo {author} {\bibfnamefont {J.}~\bibnamefont
  {Wu}},\ }\bibfield  {title} {\bibinfo {title} {Acoustical {{Tweezers}}},\
  }\href@noop {} {\bibfield  {journal} {\bibinfo  {journal} {ASA}\ }\textbf
  {\bibinfo {volume} {89}} (\bibinfo {year} {1991})}\BibitemShut {NoStop}%
\bibitem [{\citenamefont {Wang}\ \emph {et~al.}(2015)\citenamefont {Wang},
  \citenamefont {Xu}, \citenamefont {Feng}, \citenamefont {Qiu},\ and\
  \citenamefont {Liu}}]{wangDextrousAcousticTrapping2015}%
  \BibitemOpen
  \bibfield  {author} {\bibinfo {author} {\bibfnamefont {T.}~\bibnamefont
  {Wang}}, \bibinfo {author} {\bibfnamefont {S.}~\bibnamefont {Xu}}, \bibinfo
  {author} {\bibfnamefont {J.}~\bibnamefont {Feng}}, \bibinfo {author}
  {\bibfnamefont {C.}~\bibnamefont {Qiu}},\ and\ \bibinfo {author}
  {\bibfnamefont {Z.}~\bibnamefont {Liu}},\ }\bibfield  {title} {\bibinfo
  {title} {Dextrous acoustic trapping and patterning of particles assisted by
  phononic crystal plate},\ }\href@noop {} {\bibfield  {journal} {\bibinfo
  {journal} {Appl Phys Lett}\ }\textbf {\bibinfo {volume} {106}} (\bibinfo
  {year} {2015})}\BibitemShut {NoStop}%
\bibitem [{\citenamefont {Abdelaziz}\ and\ \citenamefont
  {Grier}(2021)}]{abdelazizDynamicsAcousticallyTrapped2021}%
  \BibitemOpen
  \bibfield  {author} {\bibinfo {author} {\bibfnamefont {M.~A.}\ \bibnamefont
  {Abdelaziz}}\ and\ \bibinfo {author} {\bibfnamefont {D.~G.}\ \bibnamefont
  {Grier}},\ }\bibfield  {title} {\bibinfo {title} {Dynamics of an acoustically
  trapped sphere in beating sound waves},\ }\href
  {https://doi.org/10.1103/PhysRevResearch.3.013079} {\bibfield  {journal}
  {\bibinfo  {journal} {Phys. Rev. Research}\ }\textbf {\bibinfo {volume}
  {3}},\ \bibinfo {pages} {013079} (\bibinfo {year} {2021})}\BibitemShut
  {NoStop}%
\bibitem [{\citenamefont {Marzo}\ \emph {et~al.}(2015)\citenamefont {Marzo},
  \citenamefont {Seah}, \citenamefont {Drinkwater}, \citenamefont {Sahoo},
  \citenamefont {Long},\ and\ \citenamefont
  {Subramanian}}]{marzoHolographicAcousticElements2015a}%
  \BibitemOpen
  \bibfield  {author} {\bibinfo {author} {\bibfnamefont {A.}~\bibnamefont
  {Marzo}}, \bibinfo {author} {\bibfnamefont {S.~A.}\ \bibnamefont {Seah}},
  \bibinfo {author} {\bibfnamefont {B.~W.}\ \bibnamefont {Drinkwater}},
  \bibinfo {author} {\bibfnamefont {D.~R.}\ \bibnamefont {Sahoo}}, \bibinfo
  {author} {\bibfnamefont {B.}~\bibnamefont {Long}},\ and\ \bibinfo {author}
  {\bibfnamefont {S.}~\bibnamefont {Subramanian}},\ }\bibfield  {title}
  {\bibinfo {title} {Holographic acoustic elements for manipulation of
  levitated objects},\ }\href@noop {} {\bibfield  {journal} {\bibinfo
  {journal} {Nature Communications}\ }\textbf {\bibinfo {volume} {6}} (\bibinfo
  {year} {2015})}\BibitemShut {NoStop}%
\bibitem [{\citenamefont {Baresch}\ \emph {et~al.}(2016)\citenamefont
  {Baresch}, \citenamefont {Thomas},\ and\ \citenamefont
  {Marchiano}}]{bareschObservationSingleBeamGradient2016}%
  \BibitemOpen
  \bibfield  {author} {\bibinfo {author} {\bibfnamefont {D.}~\bibnamefont
  {Baresch}}, \bibinfo {author} {\bibfnamefont {J.-L.}\ \bibnamefont
  {Thomas}},\ and\ \bibinfo {author} {\bibfnamefont {R.}~\bibnamefont
  {Marchiano}},\ }\bibfield  {title} {\bibinfo {title} {Observation of a
  {{Single}}-{{Beam Gradient Force Acoustical Trap}} for {{Elastic Particles}}:
  Acoustical {{Tweezers}}},\ }\href@noop {} {\bibfield  {journal} {\bibinfo
  {journal} {Phys. Rev. Lett.}\ }\textbf {\bibinfo {volume} {116}} (\bibinfo
  {year} {2016})}\BibitemShut {NoStop}%
\bibitem [{\citenamefont {Lee}\ \emph {et~al.}(2009)\citenamefont {Lee},
  \citenamefont {Teh}, \citenamefont {Lee}, \citenamefont {Kim}, \citenamefont
  {Lee},\ and\ \citenamefont {Shung}}]{leeSingleBeamAcoustic2009}%
  \BibitemOpen
  \bibfield  {author} {\bibinfo {author} {\bibfnamefont {J.}~\bibnamefont
  {Lee}}, \bibinfo {author} {\bibfnamefont {S.-Y.}\ \bibnamefont {Teh}},
  \bibinfo {author} {\bibfnamefont {A.}~\bibnamefont {Lee}}, \bibinfo {author}
  {\bibfnamefont {H.~H.}\ \bibnamefont {Kim}}, \bibinfo {author} {\bibfnamefont
  {C.}~\bibnamefont {Lee}},\ and\ \bibinfo {author} {\bibfnamefont {K.~K.}\
  \bibnamefont {Shung}},\ }\bibfield  {title} {\bibinfo {title} {Single {{Beam
  Acoustic Trapping}}},\ }\href@noop {} {\bibfield  {journal} {\bibinfo
  {journal} {Appl Phys Lett}\ }\textbf {\bibinfo {volume} {95}} (\bibinfo
  {year} {2009})}\BibitemShut {NoStop}%
\bibitem [{\citenamefont {Urbansky}\ \emph {et~al.}(2017)\citenamefont
  {Urbansky}, \citenamefont {Ohlsson}, \citenamefont {Lenshof}, \citenamefont
  {Garofalo}, \citenamefont {Scheding},\ and\ \citenamefont
  {Laurell}}]{urbanskyRapidEffectiveEnrichment2017}%
  \BibitemOpen
  \bibfield  {author} {\bibinfo {author} {\bibfnamefont {A.}~\bibnamefont
  {Urbansky}}, \bibinfo {author} {\bibfnamefont {P.}~\bibnamefont {Ohlsson}},
  \bibinfo {author} {\bibfnamefont {A.}~\bibnamefont {Lenshof}}, \bibinfo
  {author} {\bibfnamefont {F.}~\bibnamefont {Garofalo}}, \bibinfo {author}
  {\bibfnamefont {S.}~\bibnamefont {Scheding}},\ and\ \bibinfo {author}
  {\bibfnamefont {T.}~\bibnamefont {Laurell}},\ }\bibfield  {title} {\bibinfo
  {title} {Rapid and effective enrichment of mononuclear cells from blood using
  acoustophoresis},\ }\href {https://doi.org/10.1038/s41598-017-17200-9}
  {\bibfield  {journal} {\bibinfo  {journal} {Sci Rep}\ }\textbf {\bibinfo
  {volume} {7}},\ \bibinfo {pages} {17161} (\bibinfo {year}
  {2017})}\BibitemShut {NoStop}%
\bibitem [{\citenamefont {J{\"o}nsson}\ \emph {et~al.}(2004)\citenamefont
  {J{\"o}nsson}, \citenamefont {Holm}, \citenamefont {Nilsson}, \citenamefont
  {Petersson}, \citenamefont {Johnsson},\ and\ \citenamefont
  {Laurell}}]{jonssonParticleSeparationUsing2004}%
  \BibitemOpen
  \bibfield  {author} {\bibinfo {author} {\bibfnamefont {H.}~\bibnamefont
  {J{\"o}nsson}}, \bibinfo {author} {\bibfnamefont {C.}~\bibnamefont {Holm}},
  \bibinfo {author} {\bibfnamefont {A.}~\bibnamefont {Nilsson}}, \bibinfo
  {author} {\bibfnamefont {F.}~\bibnamefont {Petersson}}, \bibinfo {author}
  {\bibfnamefont {P.}~\bibnamefont {Johnsson}},\ and\ \bibinfo {author}
  {\bibfnamefont {T.}~\bibnamefont {Laurell}},\ }\bibfield  {title} {\bibinfo
  {title} {Particle {{Separation Using Ultrasound Can Radically Reduce Embolic
  Load}} to {{Brain After Cardiac Surgery}}},\ }\href
  {https://doi.org/10.1016/j.athoracsur.2004.04.071} {\bibfield  {journal}
  {\bibinfo  {journal} {The Annals of Thoracic Surgery}\ }\textbf {\bibinfo
  {volume} {78}},\ \bibinfo {pages} {1572} (\bibinfo {year}
  {2004})}\BibitemShut {NoStop}%
\bibitem [{\citenamefont {Wang}\ \emph {et~al.}(2017)\citenamefont {Wang},
  \citenamefont {Qiu}, \citenamefont {Zhang}, \citenamefont {Han},
  \citenamefont {Ke},\ and\ \citenamefont
  {Liu}}]{wangSoundmediatedStableConfigurations2017a}%
  \BibitemOpen
  \bibfield  {author} {\bibinfo {author} {\bibfnamefont {M.}~\bibnamefont
  {Wang}}, \bibinfo {author} {\bibfnamefont {C.}~\bibnamefont {Qiu}}, \bibinfo
  {author} {\bibfnamefont {S.}~\bibnamefont {Zhang}}, \bibinfo {author}
  {\bibfnamefont {R.}~\bibnamefont {Han}}, \bibinfo {author} {\bibfnamefont
  {M.}~\bibnamefont {Ke}},\ and\ \bibinfo {author} {\bibfnamefont
  {Z.}~\bibnamefont {Liu}},\ }\bibfield  {title} {\bibinfo {title}
  {Sound-mediated stable configurations for polystyrene particles},\ }\href
  {https://doi.org/10.1103/PhysRevE.96.052604} {\bibfield  {journal} {\bibinfo
  {journal} {Phys. Rev. E}\ }\textbf {\bibinfo {volume} {96}},\ \bibinfo
  {pages} {052604} (\bibinfo {year} {2017})}\BibitemShut {NoStop}%
\bibitem [{\citenamefont {Lopes}\ \emph {et~al.}(2016)\citenamefont {Lopes},
  \citenamefont {Azarpeyvand},\ and\ \citenamefont
  {Silva}}]{lopesAcousticInteractionForces2016}%
  \BibitemOpen
  \bibfield  {author} {\bibinfo {author} {\bibfnamefont {J.~H.}\ \bibnamefont
  {Lopes}}, \bibinfo {author} {\bibfnamefont {M.}~\bibnamefont {Azarpeyvand}},\
  and\ \bibinfo {author} {\bibfnamefont {G.~T.}\ \bibnamefont {Silva}},\
  }\bibfield  {title} {\bibinfo {title} {Acoustic {{Interaction Forces}} and
  {{Torques Acting}} on {{Suspended Spheres}} in an {{Ideal Fluid}}},\ }\href
  {https://doi.org/10.1109/TUFFC.2015.2494693} {\bibfield  {journal} {\bibinfo
  {journal} {IEEE Trans. Ultrason., Ferroelect., Freq. Contr.}\ }\textbf
  {\bibinfo {volume} {63}},\ \bibinfo {pages} {186} (\bibinfo {year}
  {2016})}\BibitemShut {NoStop}%
\bibitem [{\citenamefont {Yoshida}\ \emph {et~al.}(2011)\citenamefont
  {Yoshida}, \citenamefont {Fujikawa},\ and\ \citenamefont
  {Watanabe}}]{yoshidaExperimentalInvestigationReversal2011}%
  \BibitemOpen
  \bibfield  {author} {\bibinfo {author} {\bibfnamefont {K.}~\bibnamefont
  {Yoshida}}, \bibinfo {author} {\bibfnamefont {T.}~\bibnamefont {Fujikawa}},\
  and\ \bibinfo {author} {\bibfnamefont {Y.}~\bibnamefont {Watanabe}},\
  }\bibfield  {title} {\bibinfo {title} {Experimental investigation on reversal
  of secondary {{Bjerknes}} force between two bubbles in ultrasonic standing
  wave},\ }\href {https://doi.org/10.1121/1.3592205} {\bibfield  {journal}
  {\bibinfo  {journal} {The Journal of the Acoustical Society of America}\
  }\textbf {\bibinfo {volume} {130}},\ \bibinfo {pages} {135} (\bibinfo {year}
  {2011})}\BibitemShut {NoStop}%
\bibitem [{\citenamefont {Castro}\ and\ \citenamefont
  {Hoyos}(2015)}]{castroDeterminationSecondaryBjerknes2015}%
  \BibitemOpen
  \bibfield  {author} {\bibinfo {author} {\bibfnamefont {L.~A.}\ \bibnamefont
  {Castro}}\ and\ \bibinfo {author} {\bibfnamefont {M.}~\bibnamefont {Hoyos}},\
  }\bibfield  {title} {\bibinfo {title} {Determination of the {{Secondary
  Bjerknes Force}} in {{Acoustic Resonators}} on {{Ground}} and in
  {{Microgravity Conditions}}},\ }\href@noop {} {\bibfield  {journal} {\bibinfo
   {journal} {Microgravity Sci. Technol.}\ }\textbf {\bibinfo {volume} {28}},\
  \bibinfo {pages} {11} (\bibinfo {year} {2015})}\BibitemShut {NoStop}%
\bibitem [{\citenamefont {Zhang}\ \emph
  {et~al.}(2016{\natexlab{a}})\citenamefont {Zhang}, \citenamefont {Zhang},\
  and\ \citenamefont {Li}}]{zhangSecondaryBjerknesForce2016}%
  \BibitemOpen
  \bibfield  {author} {\bibinfo {author} {\bibfnamefont {Y.}~\bibnamefont
  {Zhang}}, \bibinfo {author} {\bibfnamefont {Y.}~\bibnamefont {Zhang}},\ and\
  \bibinfo {author} {\bibfnamefont {S.}~\bibnamefont {Li}},\ }\bibfield
  {title} {\bibinfo {title} {The secondary {{Bjerknes}} force between two gas
  bubbles under dual-frequency acoustic excitation},\ }\href
  {https://doi.org/10.1016/j.ultsonch.2015.08.022} {\bibfield  {journal}
  {\bibinfo  {journal} {Ultrasonics Sonochemistry}\ }\textbf {\bibinfo {volume}
  {29}},\ \bibinfo {pages} {129} (\bibinfo {year}
  {2016}{\natexlab{a}})}\BibitemShut {NoStop}%
\bibitem [{\citenamefont {{Garcia-Sabat{\'e}}}\ \emph
  {et~al.}(2014)\citenamefont {{Garcia-Sabat{\'e}}}, \citenamefont {Castro},
  \citenamefont {Hoyos},\ and\ \citenamefont
  {{Gonz{\'a}lez-Cinca}}}]{garcia-sabateExperimentalStudyInterparticle2014}%
  \BibitemOpen
  \bibfield  {author} {\bibinfo {author} {\bibfnamefont {A.}~\bibnamefont
  {{Garcia-Sabat{\'e}}}}, \bibinfo {author} {\bibfnamefont {A.}~\bibnamefont
  {Castro}}, \bibinfo {author} {\bibfnamefont {M.}~\bibnamefont {Hoyos}},\ and\
  \bibinfo {author} {\bibfnamefont {R.}~\bibnamefont {{Gonz{\'a}lez-Cinca}}},\
  }\bibfield  {title} {\bibinfo {title} {Experimental study on inter-particle
  acoustic forces},\ }\href {https://doi.org/10.1121/1.4864483} {\bibfield
  {journal} {\bibinfo  {journal} {The Journal of the Acoustical Society of
  America}\ }\textbf {\bibinfo {volume} {135}},\ \bibinfo {pages} {1056}
  (\bibinfo {year} {2014})}\BibitemShut {NoStop}%
\bibitem [{\citenamefont {Baasch}\ \emph {et~al.}(2017)\citenamefont {Baasch},
  \citenamefont {Leibacher},\ and\ \citenamefont
  {Dual}}]{baaschMultibodyDynamicsAcoustophoresis2017}%
  \BibitemOpen
  \bibfield  {author} {\bibinfo {author} {\bibfnamefont {T.}~\bibnamefont
  {Baasch}}, \bibinfo {author} {\bibfnamefont {I.}~\bibnamefont {Leibacher}},\
  and\ \bibinfo {author} {\bibfnamefont {J.}~\bibnamefont {Dual}},\ }\bibfield
  {title} {\bibinfo {title} {Multibody dynamics in acoustophoresis},\ }\href
  {https://doi.org/10.1121/1.4977030} {\bibfield  {journal} {\bibinfo
  {journal} {J. Acoust. Soc. Am.}\ }\textbf {\bibinfo {volume} {141}},\
  \bibinfo {pages} {1664} (\bibinfo {year} {2017})},\ \bibinfo {note} {doesn't
  explore the Rayleigh Regime}\BibitemShut {NoStop}%
\bibitem [{\citenamefont {Silva}\ and\ \citenamefont
  {Bruus}(2014)}]{silvaAcousticInteractionForces2014}%
  \BibitemOpen
  \bibfield  {author} {\bibinfo {author} {\bibfnamefont {G.~T.}\ \bibnamefont
  {Silva}}\ and\ \bibinfo {author} {\bibfnamefont {H.}~\bibnamefont {Bruus}},\
  }\bibfield  {title} {\bibinfo {title} {Acoustic interaction forces between
  small particles in an ideal fluid},\ }\href
  {https://doi.org/10.1103/PhysRevE.90.063007} {\bibfield  {journal} {\bibinfo
  {journal} {Phys. Rev. E}\ }\textbf {\bibinfo {volume} {90}},\ \bibinfo
  {pages} {063007} (\bibinfo {year} {2014})}\BibitemShut {NoStop}%
\bibitem [{\citenamefont {{Glynne-Jones}}\ \emph {et~al.}(2013)\citenamefont
  {{Glynne-Jones}}, \citenamefont {Mishra}, \citenamefont {Boltryk},\ and\
  \citenamefont {Hill}}]{glynne-jonesEfficientFiniteElement2013}%
  \BibitemOpen
  \bibfield  {author} {\bibinfo {author} {\bibfnamefont {P.}~\bibnamefont
  {{Glynne-Jones}}}, \bibinfo {author} {\bibfnamefont {P.~P.}\ \bibnamefont
  {Mishra}}, \bibinfo {author} {\bibfnamefont {R.~J.}\ \bibnamefont
  {Boltryk}},\ and\ \bibinfo {author} {\bibfnamefont {M.}~\bibnamefont
  {Hill}},\ }\bibfield  {title} {\bibinfo {title} {Efficient finite element
  modeling of radiation forces on elastic particles of arbitrary size and
  geometry},\ }\href {https://doi.org/10.1121/1.4794393} {\bibfield  {journal}
  {\bibinfo  {journal} {The Journal of the Acoustical Society of America}\
  }\textbf {\bibinfo {volume} {133}},\ \bibinfo {pages} {1885} (\bibinfo {year}
  {2013})}\BibitemShut {NoStop}%
\bibitem [{\citenamefont {Zhang}\ \emph
  {et~al.}(2016{\natexlab{b}})\citenamefont {Zhang}, \citenamefont {Qiu},
  \citenamefont {Wang}, \citenamefont {Ke},\ and\ \citenamefont
  {Liu}}]{zhangAcousticallyMediatedLongrange2016}%
  \BibitemOpen
  \bibfield  {author} {\bibinfo {author} {\bibfnamefont {S.}~\bibnamefont
  {Zhang}}, \bibinfo {author} {\bibfnamefont {C.}~\bibnamefont {Qiu}}, \bibinfo
  {author} {\bibfnamefont {M.}~\bibnamefont {Wang}}, \bibinfo {author}
  {\bibfnamefont {M.}~\bibnamefont {Ke}},\ and\ \bibinfo {author}
  {\bibfnamefont {Z.}~\bibnamefont {Liu}},\ }\bibfield  {title} {\bibinfo
  {title} {Acoustically mediated long-range interaction among multiple
  spherical particles exposed to a plane standing wave},\ }\href
  {https://doi.org/10.1088/1367-2630/18/11/113034} {\bibfield  {journal}
  {\bibinfo  {journal} {New J. Phys.}\ }\textbf {\bibinfo {volume} {18}},\
  \bibinfo {pages} {113034} (\bibinfo {year} {2016}{\natexlab{b}})}\BibitemShut
  {NoStop}%
\bibitem [{\citenamefont {Fairweather}\ \emph {et~al.}(2003)\citenamefont
  {Fairweather}, \citenamefont {Karageorghis},\ and\ \citenamefont
  {Martin}}]{fairweatherMethodFundamentalSolutions2003a}%
  \BibitemOpen
  \bibfield  {author} {\bibinfo {author} {\bibfnamefont {G.}~\bibnamefont
  {Fairweather}}, \bibinfo {author} {\bibfnamefont {A.}~\bibnamefont
  {Karageorghis}},\ and\ \bibinfo {author} {\bibfnamefont {P.}~\bibnamefont
  {Martin}},\ }\bibfield  {title} {\bibinfo {title} {The method of fundamental
  solutions for scattering and radiation problems},\ }\href
  {https://doi.org/10.1016/S0955-7997(03)00017-1} {\bibfield  {journal}
  {\bibinfo  {journal} {Engineering Analysis with Boundary Elements}\ }\textbf
  {\bibinfo {volume} {27}},\ \bibinfo {pages} {759} (\bibinfo {year}
  {2003})}\BibitemShut {NoStop}%
\bibitem [{\citenamefont {Ramaswamy}(2017)}]{ramaswamyActiveMatter2017}%
  \BibitemOpen
  \bibfield  {author} {\bibinfo {author} {\bibfnamefont {S.}~\bibnamefont
  {Ramaswamy}},\ }\bibfield  {title} {\bibinfo {title} {Active matter},\ }\href
  {https://doi.org/10.1088/1742-5468/aa6bc5} {\bibfield  {journal} {\bibinfo
  {journal} {J. Stat. Mech.}\ }\textbf {\bibinfo {volume} {2017}},\ \bibinfo
  {pages} {054002} (\bibinfo {year} {2017})}\BibitemShut {NoStop}%
\bibitem [{\citenamefont {Kang}\ \emph {et~al.}(2019)\citenamefont {Kang},
  \citenamefont {Jenkins},\ and\ \citenamefont
  {Werner}}]{kangRecentProgressActive2019}%
  \BibitemOpen
  \bibfield  {author} {\bibinfo {author} {\bibfnamefont {L.}~\bibnamefont
  {Kang}}, \bibinfo {author} {\bibfnamefont {R.~P.}\ \bibnamefont {Jenkins}},\
  and\ \bibinfo {author} {\bibfnamefont {D.~H.}\ \bibnamefont {Werner}},\
  }\bibfield  {title} {\bibinfo {title} {Recent {{Progress}} in {{Active
  Optical Metasurfaces}}},\ }\href {https://doi.org/10.1002/adom.201801813}
  {\bibfield  {journal} {\bibinfo  {journal} {Adv. Optical Mater.}\ }\textbf
  {\bibinfo {volume} {7}},\ \bibinfo {pages} {1801813} (\bibinfo {year}
  {2019})}\BibitemShut {NoStop}%
\bibitem [{\citenamefont {Zhou}\ \emph {et~al.}(2017)\citenamefont {Zhou},
  \citenamefont {Zhao}, \citenamefont {Wei},\ and\ \citenamefont
  {Wang}}]{zhouTwistsTurnsOrbiting2017a}%
  \BibitemOpen
  \bibfield  {author} {\bibinfo {author} {\bibfnamefont {C.}~\bibnamefont
  {Zhou}}, \bibinfo {author} {\bibfnamefont {L.}~\bibnamefont {Zhao}}, \bibinfo
  {author} {\bibfnamefont {M.}~\bibnamefont {Wei}},\ and\ \bibinfo {author}
  {\bibfnamefont {W.}~\bibnamefont {Wang}},\ }\bibfield  {title} {\bibinfo
  {title} {Twists and {{Turns}} of {{Orbiting}} and {{Spinning Metallic
  Microparticles Powered}} by {{Megahertz Ultrasound}}},\ }\href
  {https://doi.org/10.1021/acsnano.7b07183} {\bibfield  {journal} {\bibinfo
  {journal} {ACS Nano}\ }\textbf {\bibinfo {volume} {11}},\ \bibinfo {pages}
  {12668} (\bibinfo {year} {2017})}\BibitemShut {NoStop}%
\bibitem [{\citenamefont {Mitri}\ \emph {et~al.}(2012)\citenamefont {Mitri},
  \citenamefont {Lobo},\ and\ \citenamefont
  {Silva}}]{mitriAxialAcousticRadiation2012}%
  \BibitemOpen
  \bibfield  {author} {\bibinfo {author} {\bibfnamefont {F.~G.}\ \bibnamefont
  {Mitri}}, \bibinfo {author} {\bibfnamefont {T.~P.}\ \bibnamefont {Lobo}},\
  and\ \bibinfo {author} {\bibfnamefont {G.~T.}\ \bibnamefont {Silva}},\
  }\bibfield  {title} {\bibinfo {title} {Axial acoustic radiation torque of a
  {{Bessel}} vortex beam on spherical shells},\ }\href
  {https://doi.org/10.1103/PhysRevE.85.026602} {\bibfield  {journal} {\bibinfo
  {journal} {Phys. Rev. E}\ }\textbf {\bibinfo {volume} {85}},\ \bibinfo
  {pages} {026602} (\bibinfo {year} {2012})}\BibitemShut {NoStop}%
\bibitem [{\citenamefont {Abdelaziz}\ \emph {et~al.}(2021)\citenamefont
  {Abdelaziz}, \citenamefont {D{\'i}az~A.}, \citenamefont {Aider},
  \citenamefont {Pine}, \citenamefont {Grier},\ and\ \citenamefont
  {Hoyos}}]{abdelazizUltrasonicChainingEmulsion2021}%
  \BibitemOpen
  \bibfield  {author} {\bibinfo {author} {\bibfnamefont {M.~A.}\ \bibnamefont
  {Abdelaziz}}, \bibinfo {author} {\bibfnamefont {J.~A.}\ \bibnamefont
  {D{\'i}az~A.}}, \bibinfo {author} {\bibfnamefont {J.-L.}\ \bibnamefont
  {Aider}}, \bibinfo {author} {\bibfnamefont {D.~J.}\ \bibnamefont {Pine}},
  \bibinfo {author} {\bibfnamefont {D.~G.}\ \bibnamefont {Grier}},\ and\
  \bibinfo {author} {\bibfnamefont {M.}~\bibnamefont {Hoyos}},\ }\bibfield
  {title} {\bibinfo {title} {Ultrasonic chaining of emulsion droplets},\ }\href
  {https://doi.org/10.1103/PhysRevResearch.3.043157} {\bibfield  {journal}
  {\bibinfo  {journal} {Phys. Rev. Research}\ }\textbf {\bibinfo {volume}
  {3}},\ \bibinfo {pages} {043157} (\bibinfo {year} {2021})}\BibitemShut
  {NoStop}%
\bibitem [{\citenamefont {Liu}\ \emph {et~al.}(2017)\citenamefont {Liu},
  \citenamefont {Yang}, \citenamefont {Ni}, \citenamefont {Guo}, \citenamefont
  {Luo}, \citenamefont {Tu}, \citenamefont {Zhang},\ and\ \citenamefont
  {Zhang}}]{liuInvestigationEffectAcoustic2017}%
  \BibitemOpen
  \bibfield  {author} {\bibinfo {author} {\bibfnamefont {S.}~\bibnamefont
  {Liu}}, \bibinfo {author} {\bibfnamefont {Y.}~\bibnamefont {Yang}}, \bibinfo
  {author} {\bibfnamefont {Z.}~\bibnamefont {Ni}}, \bibinfo {author}
  {\bibfnamefont {X.}~\bibnamefont {Guo}}, \bibinfo {author} {\bibfnamefont
  {L.}~\bibnamefont {Luo}}, \bibinfo {author} {\bibfnamefont {J.}~\bibnamefont
  {Tu}}, \bibinfo {author} {\bibfnamefont {D.}~\bibnamefont {Zhang}},\ and\
  \bibinfo {author} {\bibfnamefont {a.~J.}\ \bibnamefont {Zhang}},\ }\bibfield
  {title} {\bibinfo {title} {Investigation into the {{Effect}} of {{Acoustic
  Radiation Force}} and {{Acoustic Streaming}} on {{Particle Patterning}} in
  {{Acoustic Standing Wave Fields}}},\ }\href
  {https://doi.org/10.3390/s17071664} {\bibfield  {journal} {\bibinfo
  {journal} {Sensors}\ }\textbf {\bibinfo {volume} {17}},\ \bibinfo {pages}
  {1664} (\bibinfo {year} {2017})}\BibitemShut {NoStop}%
\bibitem [{\citenamefont {Wu}(2018)}]{wuAcousticStreamingIts2018}%
  \BibitemOpen
  \bibfield  {author} {\bibinfo {author} {\bibfnamefont {J.}~\bibnamefont
  {Wu}},\ }\bibfield  {title} {\bibinfo {title} {Acoustic {{Streaming}} and
  {{Its Applications}}},\ }\href {https://doi.org/10.3390/fluids3040108}
  {\bibfield  {journal} {\bibinfo  {journal} {Fluids}\ }\textbf {\bibinfo
  {volume} {3}},\ \bibinfo {pages} {108} (\bibinfo {year} {2018})}\BibitemShut
  {NoStop}%
\bibitem [{\citenamefont {Wiklund}\ \emph {et~al.}(2012)\citenamefont
  {Wiklund}, \citenamefont {Green},\ and\ \citenamefont
  {Ohlin}}]{wiklundAcoustofluidics14Applications2012}%
  \BibitemOpen
  \bibfield  {author} {\bibinfo {author} {\bibfnamefont {M.}~\bibnamefont
  {Wiklund}}, \bibinfo {author} {\bibfnamefont {R.}~\bibnamefont {Green}},\
  and\ \bibinfo {author} {\bibfnamefont {M.}~\bibnamefont {Ohlin}},\ }\bibfield
   {title} {\bibinfo {title} {Acoustofluidics 14: {{Applications}} of acoustic
  streaming in microfluidic devices},\ }\href
  {https://doi.org/10.1039/c2lc40203c} {\bibfield  {journal} {\bibinfo
  {journal} {Lab Chip}\ }\textbf {\bibinfo {volume} {12}},\ \bibinfo {pages}
  {2438} (\bibinfo {year} {2012})}\BibitemShut {NoStop}%
\bibitem [{\citenamefont
  {Bruus}(2012{\natexlab{b}})}]{bruusAcoustofluidics10Scaling2012a}%
  \BibitemOpen
  \bibfield  {author} {\bibinfo {author} {\bibfnamefont {H.}~\bibnamefont
  {Bruus}},\ }\bibfield  {title} {\bibinfo {title} {Acoustofluidics 10:
  {{Scaling}} laws in acoustophoresis},\ }\href
  {https://doi.org/10.1039/c2lc21261g} {\bibfield  {journal} {\bibinfo
  {journal} {Lab Chip}\ }\textbf {\bibinfo {volume} {12}},\ \bibinfo {pages}
  {1578} (\bibinfo {year} {2012}{\natexlab{b}})}\BibitemShut {NoStop}%
\bibitem [{\citenamefont {Silva}\ \emph {et~al.}(2012)\citenamefont {Silva},
  \citenamefont {Lobo},\ and\ \citenamefont
  {Mitri}}]{silvaRadiationTorqueProduced2012}%
  \BibitemOpen
  \bibfield  {author} {\bibinfo {author} {\bibfnamefont {G.~T.}\ \bibnamefont
  {Silva}}, \bibinfo {author} {\bibfnamefont {T.~P.}\ \bibnamefont {Lobo}},\
  and\ \bibinfo {author} {\bibfnamefont {F.~G.}\ \bibnamefont {Mitri}},\
  }\bibfield  {title} {\bibinfo {title} {Radiation torque produced by an
  arbitrary acoustic wave},\ }\href
  {https://doi.org/10.1209/0295-5075/97/54003} {\bibfield  {journal} {\bibinfo
  {journal} {EPL}\ }\textbf {\bibinfo {volume} {97}},\ \bibinfo {pages} {54003}
  (\bibinfo {year} {2012})}\BibitemShut {NoStop}%
\bibitem [{\citenamefont
  {Bruus}(2012{\natexlab{c}})}]{bruusAcoustofluidicsPerturbationTheory2012}%
  \BibitemOpen
  \bibfield  {author} {\bibinfo {author} {\bibfnamefont {H.}~\bibnamefont
  {Bruus}},\ }\bibfield  {title} {\bibinfo {title} {Acoustofluidics 2:
  Perturbation theory and ultrasound resonance modes},\ }\href
  {https://doi.org/10.1039/C1LC20770A} {\bibfield  {journal} {\bibinfo
  {journal} {Lab Chip}\ }\textbf {\bibinfo {volume} {12}},\ \bibinfo {pages}
  {20} (\bibinfo {year} {2012}{\natexlab{c}})}\BibitemShut {NoStop}%
\bibitem [{\citenamefont {Yurkin}\ \emph {et~al.}(2007)\citenamefont {Yurkin},
  \citenamefont {Maltsev},\ and\ \citenamefont
  {Hoekstra}}]{yurkinDiscreteDipoleApproximation2007a}%
  \BibitemOpen
  \bibfield  {author} {\bibinfo {author} {\bibfnamefont {M.}~\bibnamefont
  {Yurkin}}, \bibinfo {author} {\bibfnamefont {V.}~\bibnamefont {Maltsev}},\
  and\ \bibinfo {author} {\bibfnamefont {A.}~\bibnamefont {Hoekstra}},\
  }\bibfield  {title} {\bibinfo {title} {The discrete dipole approximation for
  simulation of light scattering by particles much larger than the
  wavelength},\ }\href {https://doi.org/10.1016/j.jqsrt.2007.01.033} {\bibfield
   {journal} {\bibinfo  {journal} {Journal of Quantitative Spectroscopy and
  Radiative Transfer}\ }\textbf {\bibinfo {volume} {106}},\ \bibinfo {pages}
  {546} (\bibinfo {year} {2007})}\BibitemShut {NoStop}%
\bibitem [{\citenamefont {Yurkin}\ and\ \citenamefont
  {Hoekstra}(2011)}]{yurkinDiscretedipoleapproximationCodeADDA2011a}%
  \BibitemOpen
  \bibfield  {author} {\bibinfo {author} {\bibfnamefont {M.~A.}\ \bibnamefont
  {Yurkin}}\ and\ \bibinfo {author} {\bibfnamefont {A.~G.}\ \bibnamefont
  {Hoekstra}},\ }\bibfield  {title} {\bibinfo {title} {The
  discrete-dipole-approximation code {{ADDA}}: Capabilities and known
  limitations},\ }\href {https://doi.org/10.1016/j.jqsrt.2011.01.031}
  {\bibfield  {journal} {\bibinfo  {journal} {Journal of Quantitative
  Spectroscopy and Radiative Transfer}\ }\textbf {\bibinfo {volume} {112}},\
  \bibinfo {pages} {2234} (\bibinfo {year} {2011})}\BibitemShut {NoStop}%
\bibitem [{\citenamefont
  {Atkinson}(1982)}]{atkinsonNumericalIntegrationSphere1982}%
  \BibitemOpen
  \bibfield  {author} {\bibinfo {author} {\bibfnamefont {K.}~\bibnamefont
  {Atkinson}},\ }\bibfield  {title} {\bibinfo {title} {Numerical
  {{Integration}} on the {{Sphere}}},\ }\href@noop {} {\bibfield  {journal}
  {\bibinfo  {journal} {J. Austral. Math. Soc.}\ }\textbf {\bibinfo {volume}
  {23}},\ \bibinfo {pages} {332} (\bibinfo {year} {1982})}\BibitemShut
  {NoStop}%
\bibitem [{\citenamefont {Li}\ \emph {et~al.}(2021)\citenamefont {Li},
  \citenamefont {Liu}, \citenamefont {Lin}, \citenamefont {Ng},\ and\
  \citenamefont {Chan}}]{liNonHermitianPhysicsOptical2021}%
  \BibitemOpen
  \bibfield  {author} {\bibinfo {author} {\bibfnamefont {X.}~\bibnamefont
  {Li}}, \bibinfo {author} {\bibfnamefont {Y.}~\bibnamefont {Liu}}, \bibinfo
  {author} {\bibfnamefont {Z.}~\bibnamefont {Lin}}, \bibinfo {author}
  {\bibfnamefont {J.}~\bibnamefont {Ng}},\ and\ \bibinfo {author}
  {\bibfnamefont {C.~T.}\ \bibnamefont {Chan}},\ }\bibfield  {title} {\bibinfo
  {title} {Non-{{Hermitian}} physics for optical manipulation uncovers inherent
  instability of large clusters},\ }\href
  {https://doi.org/10.1038/s41467-021-26732-8} {\bibfield  {journal} {\bibinfo
  {journal} {Nat Commun}\ }\textbf {\bibinfo {volume} {12}},\ \bibinfo {pages}
  {6597} (\bibinfo {year} {2021})}\BibitemShut {NoStop}%
\bibitem [{\citenamefont {Sanchez}\ \emph {et~al.}(2012)\citenamefont
  {Sanchez}, \citenamefont {Chen}, \citenamefont {DeCamp}, \citenamefont
  {Heymann},\ and\ \citenamefont
  {Dogic}}]{sanchezSpontaneousMotionHierarchically2012}%
  \BibitemOpen
  \bibfield  {author} {\bibinfo {author} {\bibfnamefont {T.}~\bibnamefont
  {Sanchez}}, \bibinfo {author} {\bibfnamefont {D.~T.~N.}\ \bibnamefont
  {Chen}}, \bibinfo {author} {\bibfnamefont {S.~J.}\ \bibnamefont {DeCamp}},
  \bibinfo {author} {\bibfnamefont {M.}~\bibnamefont {Heymann}},\ and\ \bibinfo
  {author} {\bibfnamefont {Z.}~\bibnamefont {Dogic}},\ }\bibfield  {title}
  {\bibinfo {title} {Spontaneous motion in hierarchically assembled active
  matter},\ }\href {https://doi.org/10.1038/nature11591} {\bibfield  {journal}
  {\bibinfo  {journal} {Nature}\ }\textbf {\bibinfo {volume} {491}},\ \bibinfo
  {pages} {431} (\bibinfo {year} {2012})}\BibitemShut {NoStop}%
\bibitem [{\citenamefont {Darve}(2000)}]{darveFastMultipoleMethod2000a}%
  \BibitemOpen
  \bibfield  {author} {\bibinfo {author} {\bibfnamefont {E.}~\bibnamefont
  {Darve}},\ }\bibfield  {title} {\bibinfo {title} {The {{Fast Multipole
  Method}}: Numerical {{Implementation}}},\ }\href
  {https://doi.org/10.1006/jcph.2000.6451} {\bibfield  {journal} {\bibinfo
  {journal} {Journal of Computational Physics}\ }\textbf {\bibinfo {volume}
  {160}},\ \bibinfo {pages} {195} (\bibinfo {year} {2000})}\BibitemShut
  {NoStop}%
\bibitem [{\citenamefont {Kumar}\ \emph {et~al.}(2019)\citenamefont {Kumar},
  \citenamefont {Coli}, \citenamefont {Dijkstra},\ and\ \citenamefont
  {Sastry}}]{kumarInverseDesignCharged2019}%
  \BibitemOpen
  \bibfield  {author} {\bibinfo {author} {\bibfnamefont {R.}~\bibnamefont
  {Kumar}}, \bibinfo {author} {\bibfnamefont {G.~M.}\ \bibnamefont {Coli}},
  \bibinfo {author} {\bibfnamefont {M.}~\bibnamefont {Dijkstra}},\ and\
  \bibinfo {author} {\bibfnamefont {S.}~\bibnamefont {Sastry}},\ }\bibfield
  {title} {\bibinfo {title} {Inverse design of charged colloidal particle
  interactions for self assembly into specified crystal structures},\ }\href
  {https://doi.org/10.1063/1.5111492} {\bibfield  {journal} {\bibinfo
  {journal} {J. Chem. Phys.}\ }\textbf {\bibinfo {volume} {151}},\ \bibinfo
  {pages} {084109} (\bibinfo {year} {2019})}\BibitemShut {NoStop}%
\bibitem [{\citenamefont {Adorf}\ \emph {et~al.}(2018)\citenamefont {Adorf},
  \citenamefont {Antonaglia}, \citenamefont {Dshemuchadse},\ and\ \citenamefont
  {Glotzer}}]{adorfInverseDesignSimple2018}%
  \BibitemOpen
  \bibfield  {author} {\bibinfo {author} {\bibfnamefont {C.~S.}\ \bibnamefont
  {Adorf}}, \bibinfo {author} {\bibfnamefont {J.}~\bibnamefont {Antonaglia}},
  \bibinfo {author} {\bibfnamefont {J.}~\bibnamefont {Dshemuchadse}},\ and\
  \bibinfo {author} {\bibfnamefont {S.~C.}\ \bibnamefont {Glotzer}},\
  }\bibfield  {title} {\bibinfo {title} {Inverse design of simple pair
  potentials for the self-assembly of complex structures},\ }\href
  {https://doi.org/10.1063/1.5063802} {\bibfield  {journal} {\bibinfo
  {journal} {J. Chem. Phys.}\ }\textbf {\bibinfo {volume} {149}},\ \bibinfo
  {pages} {204102} (\bibinfo {year} {2018})}\BibitemShut {NoStop}%
\bibitem [{\citenamefont {Lindquist}\ \emph {et~al.}(2016)\citenamefont
  {Lindquist}, \citenamefont {Jadrich},\ and\ \citenamefont
  {Truskett}}]{lindquistCommunicationInverseDesign2016b}%
  \BibitemOpen
  \bibfield  {author} {\bibinfo {author} {\bibfnamefont {B.~A.}\ \bibnamefont
  {Lindquist}}, \bibinfo {author} {\bibfnamefont {R.~B.}\ \bibnamefont
  {Jadrich}},\ and\ \bibinfo {author} {\bibfnamefont {T.~M.}\ \bibnamefont
  {Truskett}},\ }\bibfield  {title} {\bibinfo {title} {Communication: Inverse
  design for self-assembly via on-the-fly optimization},\ }\href
  {https://doi.org/10.1063/1.4962754} {\bibfield  {journal} {\bibinfo
  {journal} {The Journal of Chemical Physics}\ }\textbf {\bibinfo {volume}
  {145}},\ \bibinfo {pages} {111101} (\bibinfo {year} {2016})}\BibitemShut
  {NoStop}%
\end{thebibliography}%

\section{Acknowledgements}
This work is supported by the National Science Foundation (Career Award 2046261 and DMS-1840265), and by the Air Force Office of Scientific Research (FA9550-21-1-0196).

\end{document}


\begin{center}
\textbf{\huge Dynamics of Acoustically Bound Particles: \\Supplemental Information}
\end{center}
   
\vspace{10mm}

\section{A. Weak scattering theory}
   
Bruus et.~al.\cite{bruusAcoustofluidicsAcousticRadiation2012b} showed that for small spherical scatterers ($ka\ll 1$), the trapping force can be formulated in terms of the gradient of an acoustic potential, which itself is defined in terms of incident field values at the location of the particle as well as relevant particle/fluid parameters \eqnref{eqn:potential}.
This arises in weak field scattering theory by invoking a multipole expansion to model the scattered field. 
In this case one can easily express the monopole and dipole terms of the scattered field in terms of the incident field. 
For sound hard bodies the acoustical contrast terms \eqnsref{eqn:f1}{eqn:f2} are both unity, simplifying much of the underlying theory.
\begin{align}
    \V F_{rad} &= - \vec{\nabla} U_{rad}\\
    U^{rad} &= V_{p} \bigg[\frac{1}{2} f_{1} \kappa_{0} \langle p_{in}^{2} \rangle - \frac{3}{4} f_{2} \rho_{0} \langle v_{in}^{2} \rangle  \bigg] \label{eqn:potential}\\
    f_{1} &= 1 - \frac{\kappa_{p}}{\kappa_{0}} \label{eqn:f1} \\
    f_{2} &= \frac{2 (\frac{\rho_{p}}{\rho_{0}} -1)}{2\frac{\rho_{p}}{\rho_{0}} + 1} \label{eqn:f2}
\end{align}
where subscripts p and 0 denote particle and medium properties respectively.
   
\begin{figure}[h]
    \centering
    \includegraphics[width = 0.5\textwidth]{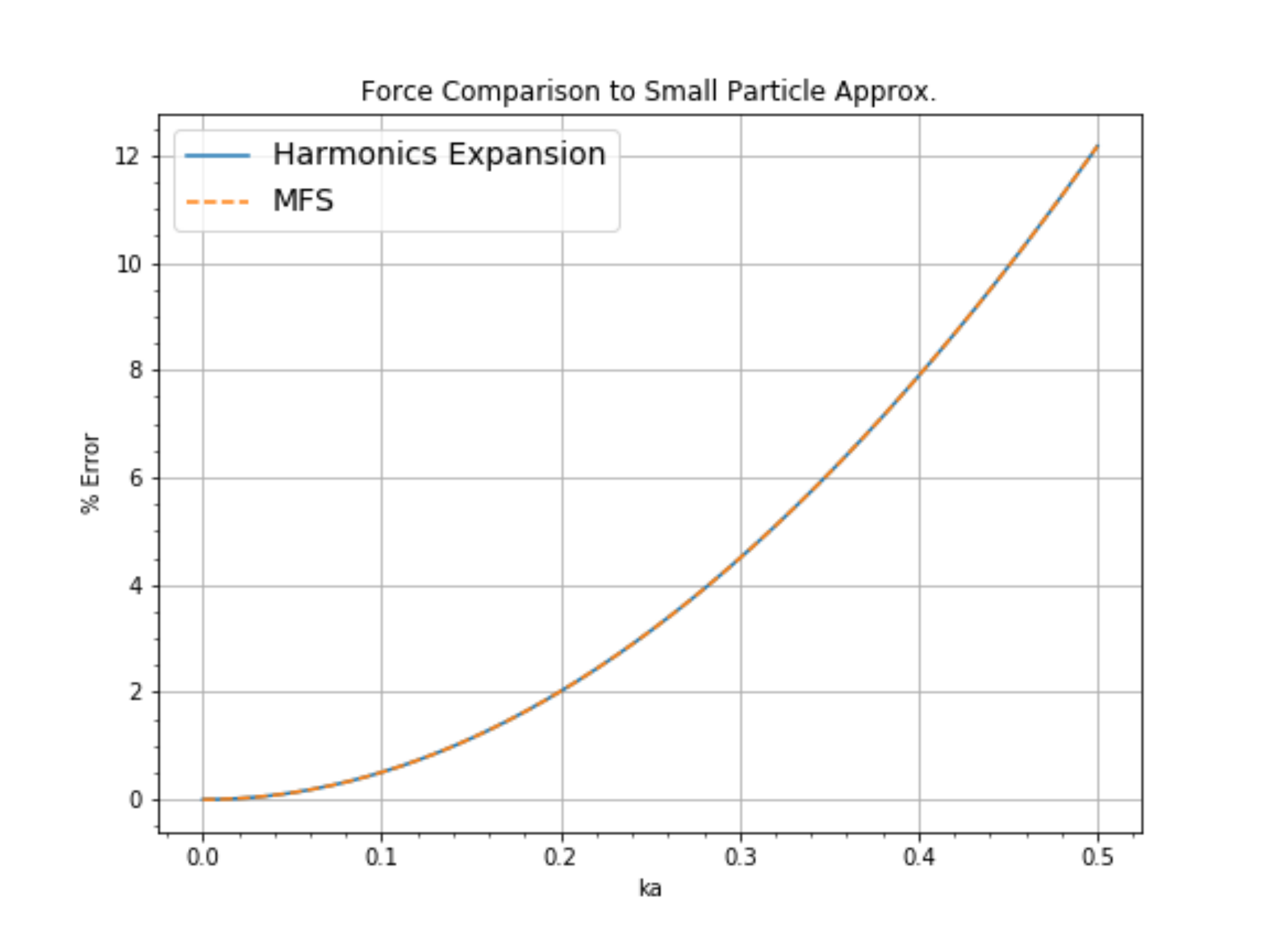}
    \caption{Percent error between acoustic trapping force computed from small particle theory and either our MFS code (dashed orange), or an exact solution (solid blue). Note how all three models give the same result as particle size approaches the Rayleigh limit.}
    \label{fig:ModelValid}
\end{figure}

We compare our MFS model to this small particle formulation in order to validate our model.
Aside from this we also compare our model to a solution given by a truncated harmonic expansion (which is exact for the case of a single particle in an axially symmetric incident field).
To compare the three models a particle is held fixed in between a node and an antinode of the external pressure wave (planar standing wave, amplitude 200Pa, frequency 40kHz), and we compute the forces according to each model as the size of the particle is varied.
We see that for $ka \ll 1$ all solutions agree, whereas for larger sized particles the weak scattering theory described by Bruus will overestimate the radiation force, while the force from the MFS model exactly matches the exact solution (\figref{fig:ModelValid}).

The exact solution can be expressed as a truncated harmonic expansion:
\begin{align}
    \phi_{sc} = \sum_{n=0}^{\infty} \sqrt{\frac{2n + 1}{4\pi}} C_{n}
    h_{n}^{(1)}(kr)P_{n}(cos\theta)
    \label{eqn:harmonics}
\end{align}
In the above $h_{n}^{(1)}$ is the spherical Hankel function of the first kind, and $P_{n}$ is the $n^{th}$ Legendre Polynomial.
The weights $C_{n}$ are chosen in order to solve the boundary condition on the surface of the sphere. For the sound hard case that boundary condition is $\partial_{r}(\phi_{in} + \phi_{sc})=0$ on the surface of the sphere.

\section{B. Numerical Methods and Model Accuracy}
We model the scattered sound field from a sound-hard spherical body using the Method of Fundamental Solutions \cite{fairweatherMethodFundamentalSolutions2003a}. 
In this method $N$ image sources are placed outside the region of interest, in this case on a concentric lattice \textit{inside} the spherical body, and the scattered field is given by a weighted superposition of fundamental solutions (Greene's functions) emanating from these image sources:
\begin{align}
    \phi_{sc} &= \sum_j^N w_j G_j(R,t) \\
    G_j(R,t) &= \frac{e^{ikR}}{R}e^{i\omega t}
\end{align}
where $k$ is the angular wavenumber, $\omega$ is the angular frequency, and $R = |\V{r}-\V{r}_j|$ is the distance from the $j^{th}$ source point to the field location $\V{r}$.
The weights are chosen by enforcing the sound-hard boundary condition ($\partial_{r}(\phi_{in} + \phi_{sc})=0$) at a second lattice of evaluation points ($\V{r}_{j'}$) on the surface of the sphere through casting the problem as a set of linear equations:
\begin{align}
    \sum_j^N \sum_{j'}^{N} \partial_r w_j G(R_{j,j'},t) &= -\partial_r \phi_1^{inc}(\V{r}_{j'}^{ bdy}) \\
    R_{j,j'} = |\V{r}_{j'}^{bdy}-\V{r}_{j}^{sca}|
\end{align}
Note that the number of source and evaluation points must be the same so that the system is exactly solvable.
Various lattice types for the source and evaluation points were tested using this model, including a fibonacci lattice, and lattices involving points subdivided onto a cube or an icosahedron and projected onto a unit sphere.
Ultimately we found that using the icosahedral projection lattice for both the source and evaluation points gave the best results due to its high degree of symmetry.
\begin{figure}[h]
    \centering
    \includegraphics[width = 0.5\textwidth]{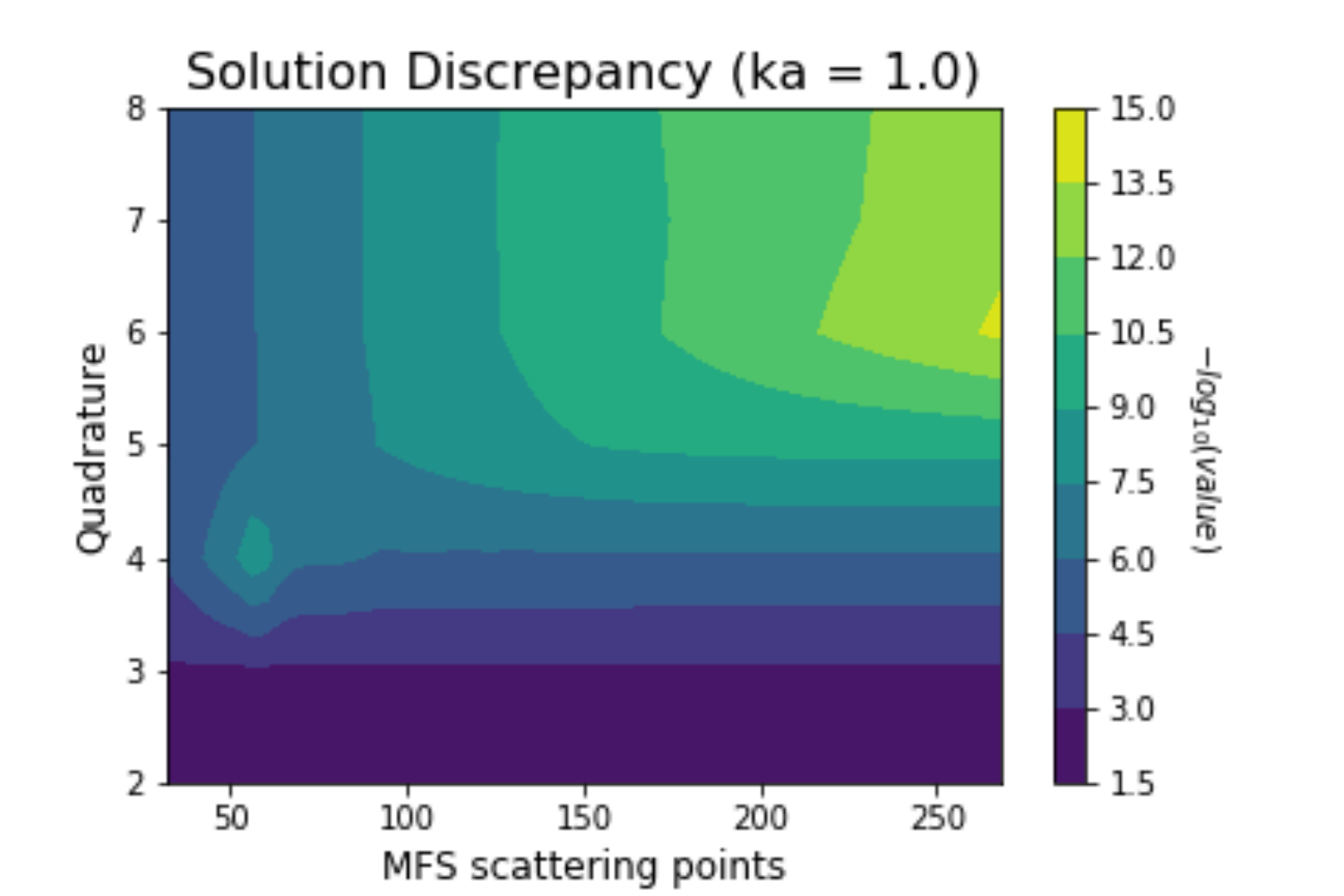}
    \caption{Discrepancy between MFS computed force and that from the Harmonic expansion. The heatmap represents the number of decimal points of error. This was done for a single particle of size $ka=1$ located between the node and antinode of the external field.
    \label{fig:Error vs source pts}
}
\end{figure}
To integrate forces over the surface of the particle, we employ the Gauss-Legendre quadrature method \cite{atkinsonNumericalIntegrationSphere1982}:
\begin{align}
    \int_{-1}^1 f(x)dx \approx \sum_i^q w_i f(x_i)
\end{align}
This method involves summing weighted values of the integrand over the surface of the sphere at various nodal points ($x_i$), in this case points spaced evenly at different (also evenly spaced) polar angles of the sphere.
The interval at each polar location is rescaled so it fits on the interval -1 to 1, and the weights ($w_i$) depend on the location of each node, as well as the legendre polynomial associated with each location on the interval from -1 to 1:
\begin{align}
    w_i &= \frac{2}{(1-x_i^2)[P'_n(x_i)]^2}
\end{align}
To determine the accuracy of our model, we again used the same simple scenario described in the section A, this time varying the number and of source points, and the number of quadrature points (upon which the Legendre-Gauss quadrature rule is applied).
\begin{figure}[h]
    \centering
    \includegraphics[width = 0.5\textwidth]{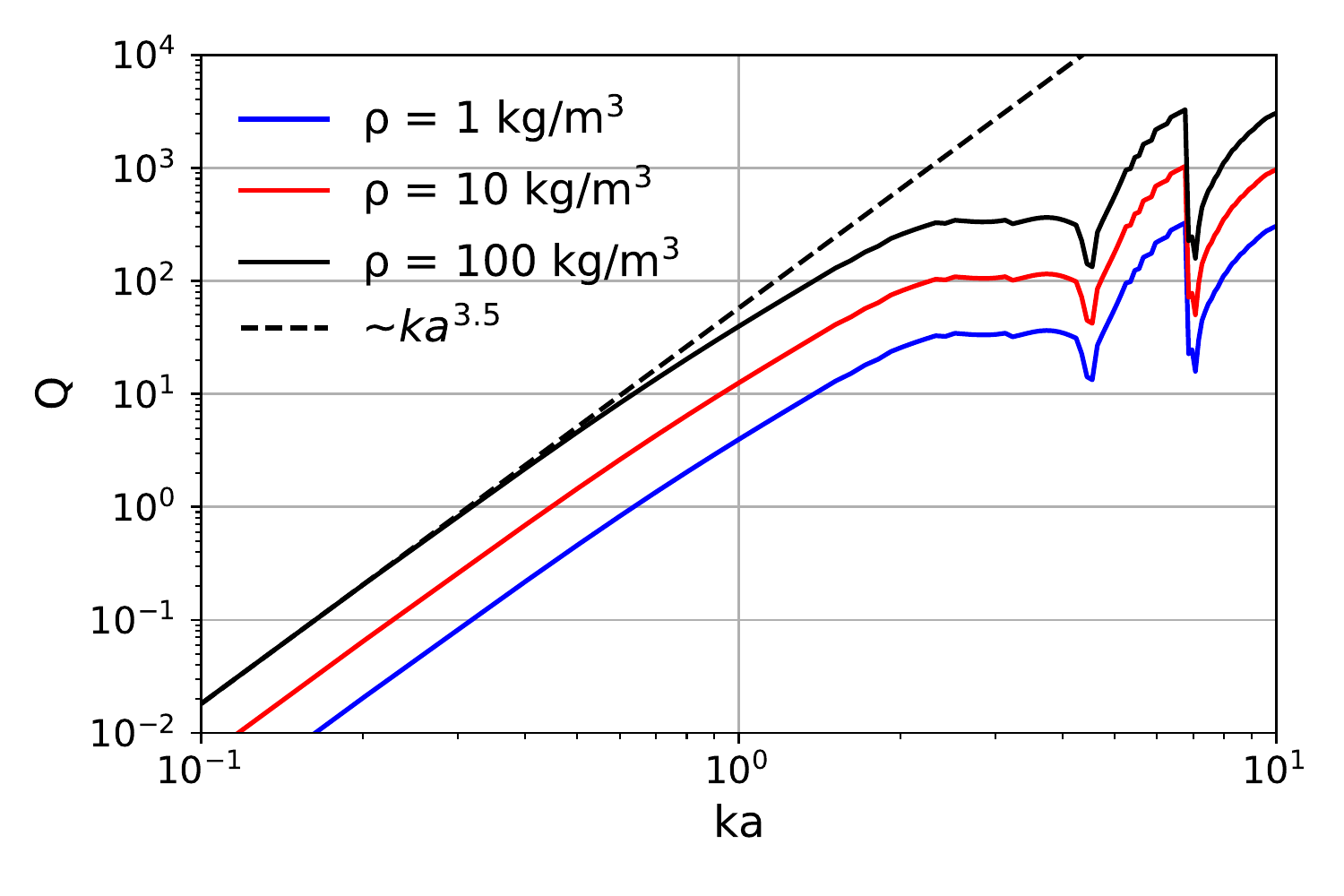}
    \caption{Quality factor vs particle size for acoustic binding between a pair of particles within a planar standing wave incident pressure field of amplitude $p_1 = 200$ Pa and frequency $\omega = 2\pi \times 40$ kHz. 
    The particles are fixed on the plane halfway between a node and an antinode and the separation distance is varied between the two particles in order to find the locations of equilibrium. 
    Effective spring constants are then taken as the slope of the binding force at these locations. 
    This plot was created using the spring constants associated with the first separation distance for the various particle densities simulated in our time-evolution studies. 
    \label{fig:quality}
}
\end{figure}
As number of source and evaluation points affects computation time, we are interested in the minimum number of points within each lattice necessary to obtain a satisfactory force computation.
To answer this question we compare the force computed using MFS to the exact solution according to the Harmonic expansion described above.
Note that the exact number of quadrature points is given by $2q^2$ where $q$ is the quadrature number given in \figref{fig:Error vs source pts}.

Python code to compute these forces, as well as notebooks used to compute errors and other quantities can be found in a Github Repository for the project \cite{klecknerlabMFSRepo2021}.

\section{C. Quality Factor}


Since we have a measure of an effective spring constant ($k_{eff}$) associated with various equilibrium locations of particles of different sizes and these bodies experience viscous linear damping, we can define the quality factor associated with these damped oscillatory systems.
\begin{align}
    Q &= \frac{\sqrt{mk_{eff}}}{6\pi \mu a} \label{eq:quality}
\end{align}
In the small particle limit the acoustic binding force (and hence associated effective spring constant) scales as the particle volume squared, using this fact together with \eqnref{eq:quality} one concludes that the quality factor scales as $ka^{7/2}$ in the Rayleigh limit (\figref{fig:quality}).



\section{D. Sound Hard Approximation}
The sound hard boundary condition describes a scenario where no bulk transmission of sound occurs in the scattering body. 
We chose this approximation as most experimental work focuses on solid bodies levitating in air, for which there is a high density and compressibility contrast between the particles and medium.
Making this approximation also simplifies the MFS solution, as we can ignore the sound field inside the particles.
Relaxing this approximation -- and allowing for sound to penetrate the scattering particles -- is possible using MFS, but requires a more complicated model which also includes an extra set of virtual scatterers.

Just how sound penetrable a body is depends on the relative measures of the density and sound speeds within a scattering body and the host medium.
For normal incidence of a planar standing wave on an infinite 2D interface, the reflection coefficient is given by \cite{meiWavePropagation2004}:
\begin{align}
    R = \frac{m-n}{m+n}\\
    m = \frac{\rho_{l}}{\rho_{u}}\\
    n = \frac{c_{u}}{c_{l}} 
\end{align}
where subscripts $u$ and $l$ denote the media above and below the refracting interface respectively.
We note that for solid bodies in a gaseous medium the speed of sound and density of the scattering body will both always be much greater than those of the gas, so that $m>1$ and $n<1$.
Even for a very light solid material  -- such as expanded polystyrene (EPS) -- the reflection coefficient is $R \sim 1$.
The density and speed of sound in air are $\rho_{air}=1.225$ kg/m$^3$ and $c_{air}=343m/s$.
The density of EPS varies, but for estimation purposed we will assume $\rho_{EPS}=10$ kg/m$^3$.
The speed of sound in EPS is dependent on the porosity, and ranges from 2000 - 3000 m/s as the porosity is decreased \cite{wangSoundmediatedStableConfigurations2017a}. 
Thus we expect a reflection coefficient of $R \sim 0.96$, implying that the sound hard approximation is appropriate in this case.

\section{E. Optical Binding Calculations}
Optical binding forces are computed using the Discrete Dipole Approximation (DDA) \cite{draineDiscreteDipoleApproximations1994,draineDiscretedipoleApproximationIts1988a}.
To implement DDA, each particle is subdivided into many individual pieces, each with a size of $\sim \lambda/10$ which is then treated like a polarizable point source.
(In practice, spherical particles are divided an exact integer number of times, so that the size of each chunk is as close to $\lambda/10$ as possible.)
The multiple scattering problem in the presence of an incoming field can then be expressed as a matrix equation, similar to MFS.
This matrix can be solved using a variety of methods \cite{yurkinDiscreteDipoleApproximation2007a}.
In our case, we use a hybrid solution where the internal scattering in a single particles is solved using an explicitly inverted internal scattering matrix, and scattering between particles is handled using an iterative approach.
In practice 3 iterations are used for $ka = 1$ -- effectively considering up to four scattering events -- and 4 iterations are used for $ka = 2-4$.
In all cases the convergence of the solution is $10^{-2}$ or better (measured as the relative root mean squared change in the dipole strength per iteration).

Forces are computed per dipole using standard techniques \cite{hoekstraRadiationForcesDiscretedipole2001b} and summed to compute per-particle forces.
We have compared the results of this solution to existing DDA simulations \cite{yurkinDiscretedipoleapproximationCodeADDA2011a} and found nearly identical results.


\bibliographystyle{apsrev4-2}
\bibliography{MFS.bib}